\documentclass[PAPER, atlasdraft=false, texlive=2023, UKenglish, cernpreprint, texmf, orcidlogo]{atlasdoc}
\usepackage[abbrevref=false]{atlaspackage}
\usepackage{atlasbiblatex}

\usepackage{atlasphysics}

\usepackage{pdflscape}
\usepackage{multirow}
\usepackage{bigstrut}
\usepackage{adjustbox}
\usepackage{PMGRefs-defs} %

\graphicspath{{logos/}{figures/}}

\usepackage{ANA-HDBS-2021-13-PAPER-defs}

\AtlasTitle{Search for a new scalar decaying
into new spin-1 bosons in four-lepton final states with the ATLAS detector}
\AtlasAbstract{%
A search is conducted for a new scalar boson~$S$, with a mass distinct from
that of the Higgs~boson, decaying promptly into four leptons ($\ell =$ $e$, $\mu$)
via an intermediate state containing two on-shell,
promptly decaying new spin-1 bosons
$Z_\text{d}$: $S \rightarrow Z_\text{d}Z_\text{d} \rightarrow 4\ell$,
where the $Z_\text{d}$ boson has a mass between 15~and 300~Ge\kern-0.1em V,
and the $S$ boson has a mass between either 30~and 115~Ge\kern-0.1em V
or 130~and 800~Ge\kern-0.1em V\@.
The search  uses proton--proton collision data
collected with the ATLAS detector at the Large Hadron Collider
with an integrated luminosity of
139~fb$^{-1}$ at a centre-of-mass energy of $\sqrt{s}=13$~Te\kern-0.1em V\@.
No significant excess above the Standard Model background expectation is observed.
Upper limits at 95\% confidence level are set on the production cross-section times branching ratio,
$\sigma(gg \to S) \times \mathcal{B}(S\rightarrow Z_\text{d}Z_\text{d} \rightarrow 4\ell)$, as a function of the mass of both particles, $m_S$ and $m_{Z\text{d}}$.

}

\author{The ATLAS Collaboration}

\AtlasRefCode{HDBS-2021-13}

\PreprintIdNumber{CERN-EP-2024-248}

\AtlasJournal{Phys.\ Lett.\ B.}
\AtlasJournalRef{\PLB 865 (2025) 139472}
\AtlasDOI{10.1016/j.physletb.2025.139472}

\hypersetup{pdftitle={ATLAS document},pdfauthor={The ATLAS Collaboration}}

\begin{document}

\maketitle

%

%

\section{Introduction}

Notwithstanding its spectacular successes, the Standard Model (SM) is
incomplete.  In particular, the nature of
dark matter is unknown.  A potential framework for extending the SM
to include dark matter is the Hidden Abelian Higgs Model
(HAHM)~\cite{Curtin:2014cca,Curtin:2013fra,Davoudiasl:2013aya,Davoudiasl:2012ag,wells2008find,gopalakrishna2008higgs},
which posits a `hidden' or `dark' sector of particles and fields,
including a field with a $U(1)_\textrm{d}$  symmetry that mixes
kinetically with the SM $U(1)_\textrm{Y}$ hypercharge gauge field
with some coupling strength~$\epsilon$.
This results in an additional
scalar~$S$ along with a new gauge boson $\Zd$ or `dark photon'.
The scalar~$S$ mixes with the SM Higgs~boson with coupling $\kappa$,
so all processes %
that can produce a SM Higgs boson also produce an $S$~boson.
Hence, the dominant production process for $S$ in proton--proton ($pp$)
collisions would be gluon--gluon fusion.
The decays of the $\Zd$~boson are determined by the gauge
couplings
and are independent of the mixings for $\epsilon,\kappa\ll 1$; the
branching ratio of the $\Zd$~boson into electron or muon pairs
would therefore vary between 10\% to 15\%
over the $\Zd$ boson mass range $1 < m_{\Zd} < \SI{300}{\GeV}$,
except in the range $60 < m_{\Zd} < \SI{135}{\GeV}$ where it is smaller,
falling to about 4\% at $m_{\Zd}=m_Z$~\cite{Curtin:2014cca}.

Previous studies~\cite{HDBS-2018-55,EXOT-2016-22,EXOT-2013-15,CMS-EXO-23-005,CMS-HIG-19-007,CMS-EXO-19-018} have searched for a resonantly produced SM Higgs boson mixing with
a new scalar decaying into a $\Zd\Zd$ or a $Z\Zd$ pair,
with each $\Zd$ or $Z$ boson decaying into a pair of electrons or muons
($\ell\equiv e, \mu$) with opposite electric charge.
Other similar searches, including searches for pairs of light bosons decaying
into muons, \tauleptons, photons, and/or jets,
as well as searches for a single light boson decaying into a pair of muons,
using both
$\sqrt{s}=\SI{8}{\tev}$ and $\sqrt{s}=\SI{13}{\tev}$ data, were performed
at the Large Hadron Collider (LHC) by the
ATLAS~\cite{HDBS-2018-47,HIGG-2017-09,HIGG-2017-05,HIGG-2016-03,HIGG-2014-02},
CMS~\cite{CMS-HIG-13-010, CMS-HIG-16-015,CMS-HIG-18-024,CMS-EXO-20-014},
and LHCb~\cite{LHCB-PAPER-2017-038} experiments.
Further searches for a SM Higgs~boson decaying into undetected particles
are reported in %
Refs.~\cite{HIGG-2021-05, CMS-HIG-20-003}. %
The present work extends the previous ATLAS $H\ra \Zd\Zd\ra 4\ell$
search~\cite{HDBS-2018-55}, where the additional scalar~$S$
decaying into the $\Zd\Zd$ pair, $S\ra \Zd\Zd\ra 4\ell$,
has a mass distinct from that of the SM Higgs~boson.
The search reported here is restricted to the region $m_S<\SI{800}{\GeV}$
and $m_{\Zd}<\SI{300}{\GeV}$.  In the HAHM, the $S$ and $\Zd$~boson
widths increase rapidly with mass, resulting in a loss of sensitivity
above these masses.
It is additionally restricted to regions of parameter space
where all decays are prompt and further makes no use of information about any possible
jets or missing transverse energy, so may also be sensitive to other
signal processes
that may produce extra particles in addition to four leptons.

%
%
%
%
%
%
%
%
%

%
%
%
%
%
%
%
%
%
%
%


\section{The ATLAS detector}

\newcommand{\AtlasCoordFootnote}{%
ATLAS uses a right-handed coordinate system with its origin at the nominal interaction point (IP)
in the centre of the detector and the \(z\)-axis along the beam pipe.
The \(x\)-axis points from the IP to the centre of the LHC ring,
and the \(y\)-axis points upwards.
Polar coordinates \((r,\phi)\) are used in the transverse plane,
\(\phi\) being the azimuthal angle around the \(z\)-axis.
The pseudorapidity is defined in terms of the polar angle \(\theta\) as \(\eta = -\ln \tan(\theta/2)\) and is equal to the rapidity
$ y = \frac{1}{2} \ln \left( \frac{E + p_z}{E - p_z} \right) $ in the relativistic limit.
Angular distance is measured in units of \(\Delta R \equiv \sqrt{(\Delta y)^{2} + (\Delta\phi)^{2}}\).}
The ATLAS experiment~\cite{PERF-2007-01} at the LHC is a multipurpose particle detector
with a forward--backward symmetric cylindrical geometry and a near \(4\pi\) coverage in
solid angle.\footnote{\AtlasCoordFootnote}
It consists of an inner tracking detector surrounded by a thin superconducting solenoid
providing a \qty{2}{\tesla} axial magnetic field, electromagnetic and hadronic calorimeters, and a muon spectrometer.
The inner tracking detector covers the pseudorapidity range \(|\eta| < 2.5\).
It consists of silicon pixel, silicon microstrip, and transition radiation tracking detectors.
Lead/liquid-argon (LAr) sampling calorimeters provide electromagnetic (EM) energy measurements
with high granularity within the region \(|\eta|< 3.2\).
A steel/scintillator-tile hadronic calorimeter covers the central pseudorapidity range (\(|\eta| < 1.7\)).
The endcap and forward regions are instrumented with LAr calorimeters
for EM and hadronic energy measurements up to \(|\eta| = 4.9\).
The muon spectrometer surrounds the calorimeters and is based on
three large superconducting air-core toroidal magnets with eight coils each.
The field integral of the toroids ranges between \num{2.0} and \qty{6.0}{\tesla\metre}
across most of the detector.
The muon spectrometer includes a system of precision tracking chambers up to \(|\eta| = 2.7\) and fast detectors for triggering up to \(|\eta| = 2.4\).
The luminosity is measured mainly by the LUCID--2~\cite{LUCID2} detector, which is located close to the beampipe.
A two-level trigger system is used to select events~\cite{TRIG-2016-01}.
The first-level trigger is implemented in hardware and uses a subset of the detector information
to accept events at a rate close to \qty{100}{\kHz}.
This is followed by a software-based trigger that
reduces the accepted rate of complete events to \qty{1.25}{\kHz} on average
depending on the data-taking conditions.
A software suite~\cite{SOFT-2022-02} is used in data simulation, in the reconstruction
and analysis of real and simulated data, in detector operations, and in the trigger and data acquisition
systems of the experiment.

\section{Data and simulated event samples}
\label{sec:samples}

The data used in this search were recorded during the 2015--2018 LHC
run with $pp$ collisions at $\sqrt{s}=\SI{13}{\tera\eV}$.
After requiring that all detector systems were
operational~\cite{DAPR-2018-01}
without excess calorimeter noise~\cite{ATLAS-CONF-2015-029},
this corresponds to
an integrated luminosity of \SI{139}{\per\femto\barn}~\cite{ATLAS-CONF-2019-021}.
Events were selected by triggers requiring either one or two
electron or muon candidates~\cite{TRIG-2018-05,TRIG-2018-01,TRIG-2019-03}.
The trigger thresholds on transverse momentum \pT range
from $\pT > \SI{7}{\GeV}$ to $\pT > \SI{60}{\GeV}$, depending
on lepton multiplicity and flavour. In either case, the trigger efficiency
is above \SI{95}{\percent} relative to the offline signal event
selection criteria.
Collision events are selected by requiring an identified primary
vertex with at least two tracks~\cite{PERF-2015-01}
and at least four lepton candidates satisfying the requirements given
in \cref{sec:event-selection}.

The expected contribution of the $S\ra \Zd\Zd\ra 4\ell$ ($\ell=e,\mu$) signal is determined
from Monte Carlo (MC) simulations.  Samples for this process
are produced according to the HAHM
implementation~\cite{hahm} for the gluon--gluon fusion production process
at leading order (LO) using \mgamc\ 2.2.2~\cite{Alwall:2014hca} with
the mixing parameters $\epsilon$ and $\kappa$ both set to $10^{-4}$
so that decays of the $\Zd$~boson are prompt~\cite{Curtin:2014cca}.
Parton showers and decays are simulated using \PYTHIA~8.186~\cite{Sjostrand:2007gs},
along with \EVTGEN 1.2.0~\cite{Lange:2001uf} to decay heavy-flavour hadrons.
The \NNPDF[2.3lo]~\cite{Ball:2012cx} set of parton distribution functions (PDFs)
and the A14 tune~\cite{ATL-PHYS-PUB-2014-021} are used.
The width of the scalar~$S$ is taken to vary with mass in the same
way as the SM Higgs~boson~\cite{Contino:2014aaa}.
The signal samples cover different  $m_S$ and $m_{\Zd}$ mass hypotheses,
in two regions.  In the first region, there are 32 samples with
$\SI{30}{\GeV} \le m_S \le \SI{110}{\GeV}$ and $\SI{15}{\GeV} \le m_{\Zd} \le \SI{48}{\GeV}$
(the upper $m_{\Zd}$ limit being chosen to avoid overlaps with the analyses
of Ref.~\cite{HDBS-2018-55}; see \cref{sec:event-selection}).
and in the second there are 49 samples with
$\SI{150}{\GeV} \le m_S \le \SI{800}{\GeV}$ and $\SI{15}{\GeV} \le m_{\Zd} \le \SI{300}{\GeV}$.
In both cases, $m_S > 2m_{\Zd}$.
Detector effects are included using a \GEANT~\cite{Agostinelli:2002hh} simulation of the ATLAS detector~\cite{SOFT-2010-01}.
The effects of additional $pp$ collisions in the same
or a neighbouring bunch crossing (pile-up) are included in the simulation.

Most background processes are estimated by using MC simulations,
which consider $\ell=e,\mu,\tau$, and thus include
leptonic $\tau$-lepton decays.  The samples
used are identical to those of Ref.~\cite{HDBS-2018-55} and are briefly
summarized here.
The non-resonant $\qqbar\to ZZ^* \to 4 \ell$ and $gg\to ZZ^* \to 4 \ell$
background processes are simulated using \sherpa~2.2.2~\cite{Bothmann:2019yzt,Gleisberg:2008fv, Cascioli:2011va, Schumann:2007mg, Hoeche:2012yf,Caola:2015psa,ATL-PHYS-PUB-2017-005}
at next-to-leading order (NLO) for up to one additional parton and at LO
for up to three additional partons.
The $H\rightarrow ZZ^*\rightarrow 4\ell$ background
is simulated with
\POWHEGBOX~v2~\cite{Hamilton:2013fea,Hamilton:2015nsa,Alioli:2010xd,Nason:2004rx,Frixione:2007vw} for the gluon--gluon fusion, vector--boson fusion,
and $\ttbar H$ processes, with \PYTHIA~8.186
for the $VH$ process, and with \mgamc for the $\bbbar H$ process.
Higher-order electroweak processes, including triboson production~($VVV$)
and vector--boson scattering~(VBS), are simulated using
\sherpa~2.2.2, while
the process $\ttbar + (Z\to \ell\ell)$ is simulated with \sherpa~2.2.0.
Other, reducible, backgrounds have fewer than four prompt leptons in the
final state, but can be accepted by the event selection if there are
additional leptons from, for example, heavy-flavour decay or jets misidentified
as leptons.  Backgrounds due to $Z+\textrm{jets}$ and $\ttbar$ processes
are estimated from data (see \cref{sec:background}), but the method also
requires simulations of these processes;
the $Z+\textrm{jets}$ process is simulated
with \sherpa~2.2.1, while the $\ttbar$ process
is simulated with \POWHEGBOX~v2.
The $WZ$ process was generated with \POWHEGBOX at NLO\@.
For processes simulated with \mgamc and \POWHEGBOX~v2,
\PYTHIA is used to simulate parton
showers and decays along with \EVTGEN.
For the portion of the $gg\to ZZ^*\to 4\ell$ sample
with $m_{4\ell}>\SI{130}{\GeV}$ (corresponding to a few percent
of the total estimated background) and the $VVV$ samples,
detector effects are included using a fast simulation~\cite{SOFT-2010-01},
of just the calorimeter
response~\cite{ATL-PHYS-PUB-2010-013} component, which relies on a parameterization.  All other samples use the full
\GEANT simulation.

In addition to the backgrounds mentioned above, numerous other simulated
background processes were checked, including $\ttbar$ associated with
a diboson pair, $VH$ with the Higgs boson decaying into two $W$ bosons,
$Z\Upsilon$, $tH$, $tWH$, and $q\bar q\rightarrow ZZ^*$ at low $\pT$.
The contributions from these backgrounds were less than
0.1\% of the total, so they were not included in the further
background calculations.

%
%
%
%
%
%
%
%
%
%
%
%
%
%
%
%
%
%
%
%
%
%
%
%
%
%
%
%
%
%


\section{Event selection}
\label{sec:event-selection}

After reconstruction, electrons and muons are identified using the
standard ATLAS `Loose' criteria~\cite{EGAM-2018-01,MUON-2018-03},
defined to maximize the reconstruction and identification efficiencies while
still providing good-quality candidates.
Electron candidates must satisfy $\pt>\SI{7}{\gev}$, $|\eta|<2.47$,
and must also have an additional associated hit in the innermost
layer of the silicon detector.
Longitudinal and transverse impact parameters with respect
to the primary vertex must satisfy
$|z_0 \sin\theta| < \SI{0.5}{\mm}$ and $|d_0| / \sigma_{d_0} < 5$.
Muon candidates must satisfy $\pt>\SI{5}{\gev}$ and $|\eta|<2.5$,
except that `calorimeter-tagged' (CT) muons,
identified by matching an inner-detector (ID) track (with no corresponding
track in the muon system) with a deposit in the calorimeter consistent
with a minimum-ionizing particle,
must instead satisfy $\pt>\SI{15}{\gev}$.
Except for `stand-alone' (SA) muons lacking a matching ID track,
muon candidates must also have impact parameters satisfying
$|z_0 \sin\theta| < \SI{0.5}{\mm}$, $|d_0| < \SI{1}{\mm}$,
and $|d_0| / \sigma_{d_0} < 3$.
Leptons must be isolated from other deposits in the
calorimeter or ID tracks~\cite{EGAM-2018-01,MUON-2018-03}.
This rejects backgrounds in which leptons arise from the decay
of heavy-flavour hadrons, or in which jets are misidentified
as leptons~\cite{EGAM-2019-01}.
For each lepton, the sum of the transverse energies
of topological clusters~\cite{PERF-2014-07}
within a cone of $\Delta R = 0.2$ around
it (excluding energy attributed to the lepton itself)
must be less than 20\% of its \pt for electrons,
and less than 30\% of its \pt for muons.
The transverse momenta of tracks in a cone around the lepton are also
summed, and must be less than 15\% of its \pt.  The radius of the
cone depends on the momentum of the lepton.  For electrons,
it is $\Delta R = \min(0.2, \SI{10}{\GeV}/\pt$), while
for muons it is $\Delta R = \min(0.3, \SI{10}{\GeV}/\pt$).

Quadruplets are then formed from two same-flavour,
opposite-sign lepton pairs: either $4e$, $2e2\mu$, or $4\mu$.
For the computation of isolation of a lepton in a quadruplet,
tracks and energy clusters attributed to the other leptons
in the same quadruplet are excluded from the sums.
A quadruplet may contain no more than one CT or SA muon.
In the case of four same-flavour leptons, two quadruplets are formed,
one for each possible pairing.  If there are more than four
leptons then there may be additional quadruplets, depending on the
flavours and signs of the leptons.

For each possible quadruplet,
the invariant masses of the two pairs are denoted $m_{ab}$ and
$m_{cd}$, where $m_{ab} > m_{cd}$.  If all four leptons
have the same flavour, then the alternative pairing $m_{ad,bc}$ can be
defined taking the positively charged lepton of the $m_{ab}$ pair
and the negatively charged lepton of the $m_{cd}$ pair to make $m_{ad}$
and the remaining two leptons to make $m_{bc}$.\footnote{These
$m_{ab,cd}$ variables are distinct from the $m_{12,34}$ variables used
in Ref.~\cite{HDBS-2018-55}, which were chosen such that
$|m_{12} - m_Z| < |m_{34}-m_Z|$.  For $m_{4\ell} < 2m_Z$,
$m_{ab,cd}=m_{12,34}$.  For larger $m_{4\ell}$, either
$m_{ab,cd}=m_{12,34}$ or $m_{ab,cd}=m_{34,12}$.}

Each quadruplet must contain all the leptons corresponding to at least
one of the triggers satisfied by the event.
The three highest-$\pt$ leptons must satisfy
$\ptn{$\ell_1$} > \SI{20}{\GeV}$, $\ptn{$\ell_2$} > \SI{15}{\GeV}$, and $\ptn{$\ell_3$} > \SI{10}{\GeV}$.
To remove poorly-measured leptons and electrons from bremsstrahlung,
all pairs of same-flavour leptons in the quadruplet
must satisfy a separation requirement of $\Delta R(\ell,\ell') > 0.1$, while different-flavour
pairs must satisfy $\Delta R(\ell,\ell') > 0.2$.
At least one quadruplet per event is required.
If more than one quadruplet
passes these requirements, the one with the smallest mass difference
between the two pairs, $\Delta m_{\ell\ell} = |m_{ab}-m_{cd}|$,
is chosen.

All dilepton pair masses $m_{ab}$, $m_{cd}$, $m_{ad}$, and $m_{bc}$ must be
larger than $\SI{11.105}{\GeV}$, corresponding to $m_{\Upsilon(3S)}+\SI{0.75}{\GeV}$,
where $m_{\Upsilon(3S)}$ is taken to be $\SI{10.355}{\GeV}$~\cite{pdg24}.
This removes events where the alternate pairing may be consistent
with $ZZ^*$ decay as well as events with
lepton pairs consistent with $J/\psi$ or $\Upsilon$ decay.

Two signal regions are defined, each considering a
scalar $S$ with a mass different from that of the SM~Higgs~boson.
The lower mass signal region (SR1) requires $m_{4\ell} < \SI{115}{\GeV}$
and, to reject background from $Z$~bosons,
either $m_{ab} < \SI{50}{\GeV}$ or $m_{ab} > \SI{106}{\GeV}$
(this is the complement of one of the signal region requirements
from the $H\rightarrow Z\Zd\rightarrow 4\ell$ search in Ref.~\cite{HDBS-2018-55},
and was chosen to avoid overlap with that search).
The higher mass signal region (SR2) requires $m_{4\ell} > \SI{130}{\GeV}$,
and $|m_{ab,cd} - m_Z| > \SI{8}{\GeV}$ and $|m_{ad,bc} - m_Z| > \SI{4}{\GeV}$,
where the latter requirement on the alternative pairing
applies only to the $4e$ and $4\mu$ events.
These pair mass requirements were set to optimize the expected
significance.

Consistency between $m_{ab}$ and $m_{cd}$ is enforced by requiring
$m_{cd}/m_{ab} > 0.85 - 0.1125f(m_{ab})$, where the modulating
function $f$ decreases monotonically
from $f(m_{ab} \le \SI{10}{\GeV})=1$ to $f(m_{ab} \ge \SI{50}{\GeV}) = 0$.
It is the result of re-optimizing the analysis to allow for larger
$\Zd$ widths at lower lepton-pair invariant masses, where the background
is low, and is fully described in Ref.~\cite{HDBS-2018-55}.

Finally, if $E'_{ab}$ is defined as the energy of the $ab$ dilepton pair
in the rest frame of the four-lepton system, then
\begin{equation*}{E'_{ab}\over m_{4\ell}} = {1\over 2}\left(
1 + {{m^2_{ab} - m^2_{cd}} \over m^2_{4\ell}} \right) .
\end{equation*}
The additional requirement $|E'_{ab}/m_{4\ell} - 0.5| < 0.008$ reduces
the dominant $ZZ^*$ background by about a factor of 1.5,
while reducing the efficiency for the signal by less than five percent
over most of the explored ($m_S, m_{\Zd}$) range.

%
%
%
%
%
%
%
%

%
%
%
%
%
%
%
%
%
%
%
%


\section{Background estimation}
\label{sec:background}

Backgrounds with four prompt leptons are estimated directly from simulation
(see \cref{sec:samples}).  The dominant background, comprising 90\%--95\%
of the total, is the non-resonant process $ZZ^* \ra 4\ell$.
As described in Ref.~\cite{HDBS-2018-55}, the prediction of this background
was verified using background-dominated validation regions.%
\footnote{In particular, Validation Region 3 of Ref.~\cite{HDBS-2018-55}
requires $m_{4\ell} < \SI{115}{\GeV}$ or $m_{4\ell} > \SI{130}{\GeV}$
and thus probes the $ZZ^*\ra4\ell$ background that is dominant
for this analysis.
}
The background from SM Higgs~boson production is effectively suppressed
by the requirement that the overall invariant mass $m_{4\ell}$ not be consistent
with that of the SM~Higgs~boson.  Other processes with four prompt
leptons include $\ttbar Z\ra 4\ell+X$
and processes with three gauge bosons and are found to be small in comparison
with the dominant $ZZ^*$ background.
The reducible background due to $WZ$ production is similarly small.

Contributions from $Z+\textrm{jets}$ and $\ttbar$ (including $\ttbar Z$
decays with fewer than four leptons)
processes to the signal regions are estimated from data.
They are estimated separately but with similar techniques.
In both cases, the event sample is enlarged by relaxing the requirements
on isolation and impact parameters for two of the lepton candidates.  For the
$\ttbar$ background, the two candidates with largest $\pT$ must satisfy the
nominal requirements.  For the $Z+\textrm{jets}$ background, the two
candidates in the pair with invariant mass closest to that of the $Z$~boson
must satisfy the nominal requirements.  In both cases, events are then
classified into four regions based on the requirements satisfied
by the other two candidates:
\begin{itemize}
\item Region A: The remaining two candidates both satisfy isolation and
impact parameter requirements.
\item Region B: The remaining two candidates both satisfy the isolation requirement,
but at least one does not satisfy the impact parameter requirement.
\item Region C: The remaining two candidates both satisfy the impact parameter
requirement, but at least one does not satisfy the isolation requirement.
\item Region D: All other events.
\end{itemize}

To obtain adequate statistics, event selections are applied only
as far as the quarkonia veto, and the electron identification is also
relaxed.  Backgrounds other than the one being estimated are subtracted
from the event counts in regions B, C, and D, using estimates from simulation.
The number of background
events in the signal region A can then be estimated as
$N_{\textrm{A}} = \epsilon_{\mathrm{sel}} N_{\textrm{B}}N_{\textrm{C}} / N_{\textrm{D}}$,
where $\epsilon_{\mathrm{sel}}$ is the efficiency of the remaining selections as determined
from simulation.  This assumes that the isolation and impact parameter
requirements are uncorrelated.  This is nearly the case for the
$4e$ and $2e2\mu$ final states (correlation coefficient $r < 0.01$),
but they are more correlated for the $4\mu$ final state ($r\approx 0.4$).
The effect of the correlation is taken into account as a systematic
uncertainty assigned to the yields of the $Z+\textrm{jets}$ and $\ttbar$
backgrounds.

%
%
%


\section{Systematic uncertainties}

The uncertainty in the integrated
luminosity is 1.7\%~\cite{ATLAS-CONF-2019-021},
obtained using the LUCID-2 detector~\cite{LUCID2}.
Uncertainties in the yields and efficiencies of simulated samples due to pile-up
arise from differences between the predicted and measured
inelastic cross-sections and from the reweighting procedure.
These uncertainties are approximately~1\%.

The efficiency for events
to satisfy the selection depends on the triggering, reconstruction, identification,
and isolation
efficiencies for leptons, as well as the determination of their momentum
scale.
Tag-and-probe techniques are applied to
dilepton resonances, including $Z\to \ell^+\ell^-$,
$J/\psi\to \ell^+\ell^-$, and $\Upsilon\to \mu^+\mu^-$,
to measure the efficiencies as well as momentum scales
and resolutions for electrons and muons.
This leads to corrections to account for differences
between data and simulation along with an estimate of the
residual uncertainty~\cite{EGAM-2018-01, MUON-2018-03}.
As there are four leptons in the final state, small single-lepton
uncertainties can result in larger uncertainties in the final
yields and efficiencies for simulated samples,
up to 17\%, dominated by the uncertainties
on the electron reconstruction and identification efficiencies.

Uncertainties in the modelling of the simulated signal and
background processes are estimated by varying the
PDFs according to the prescription of Ref.~\cite{Butterworth:2015oua}
and the factorization and renormalization scales up and down by a factor of two.
For the signal process, the resulting yield uncertainties are taken from
the results for gluon--gluon fusion production of Ref.~\cite{deFlorian:2016spz}
and vary from 10\%--19\% depending on $m_S$,
dominated by the uncertainty from factorization and renormalization scale variations.
For the dominant $ZZ^*\to 4\ell$ background, these same sources result
in an uncertainty in the yield of about 6\%.
Additionally, for this sample, the matrix element matching scale is varied
from the nominal value of \SI{20}{\GeV} to \SI{15}{\GeV} and \SI{30}{\GeV};
the resummation scale is varied up and down by a factor of four;
and the alternate recoil scheme of Ref.~\cite{Hoeche:2009xc}
(\SHERPA parameter \verb|CSS_KIN_SCHEME=1|) is compared.
This results in an uncertainty in the yield of~6\% in~SR1
and~12\% in~SR2, dominated by the matrix element matching scale.
The uncertainty in the yield for the $H\to ZZ^*\to 4\ell$ process
is about~9\%~\cite{LHCHiggsCrossSectionWorkingGroup:2011ti}.
The uncertainty in the data-driven $Z+\textrm{jets} / \ttbar$
background estimate is 50\%--100\%.

%
%
%
%
%
%
%
%
%


\section{Results}

\begin{figure}[!tbp]
\begin{center}
\subfloat[]{\includegraphics[width=0.49\textwidth]{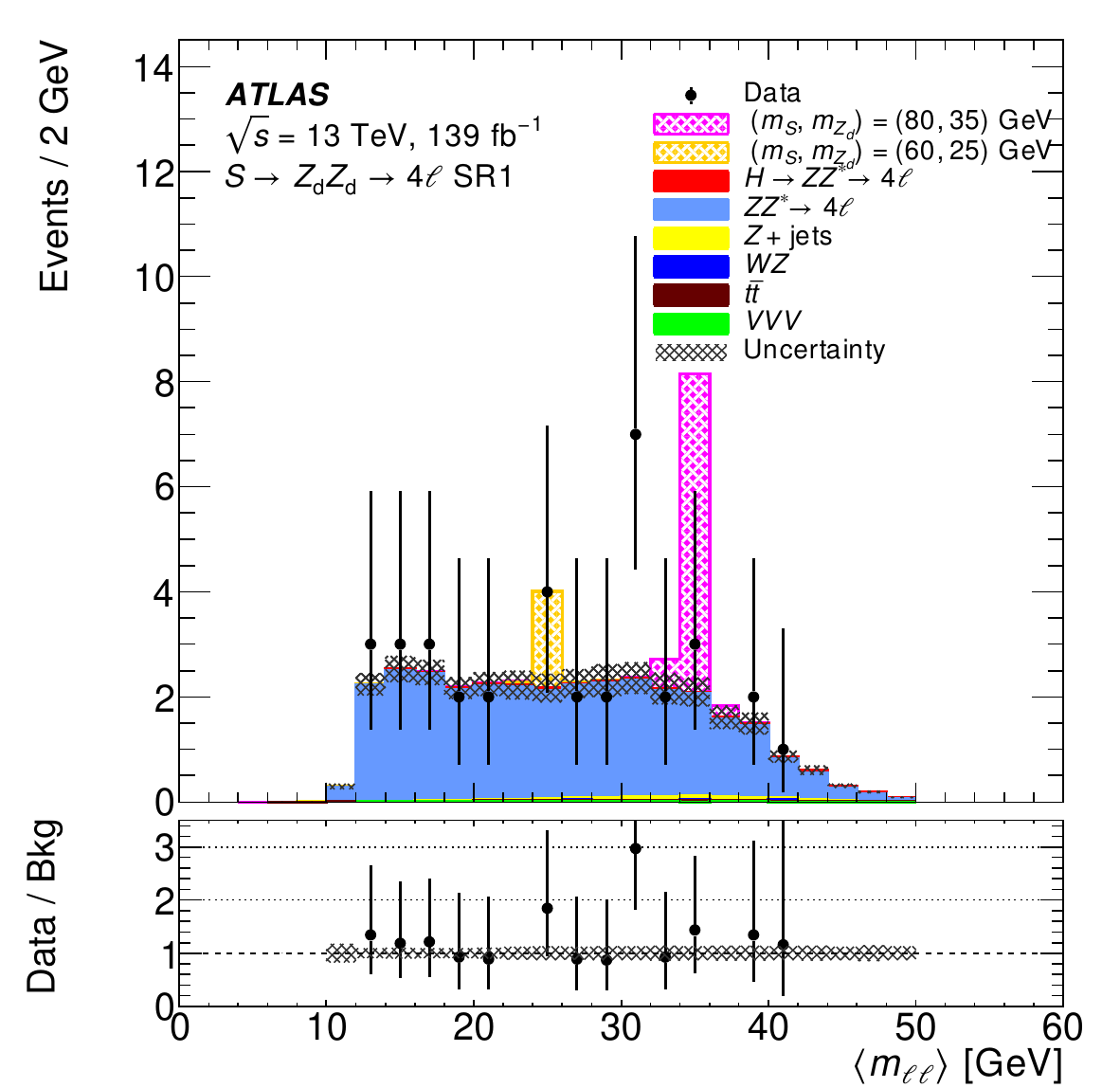}%
}
\subfloat[]{\includegraphics[width=0.49\textwidth]{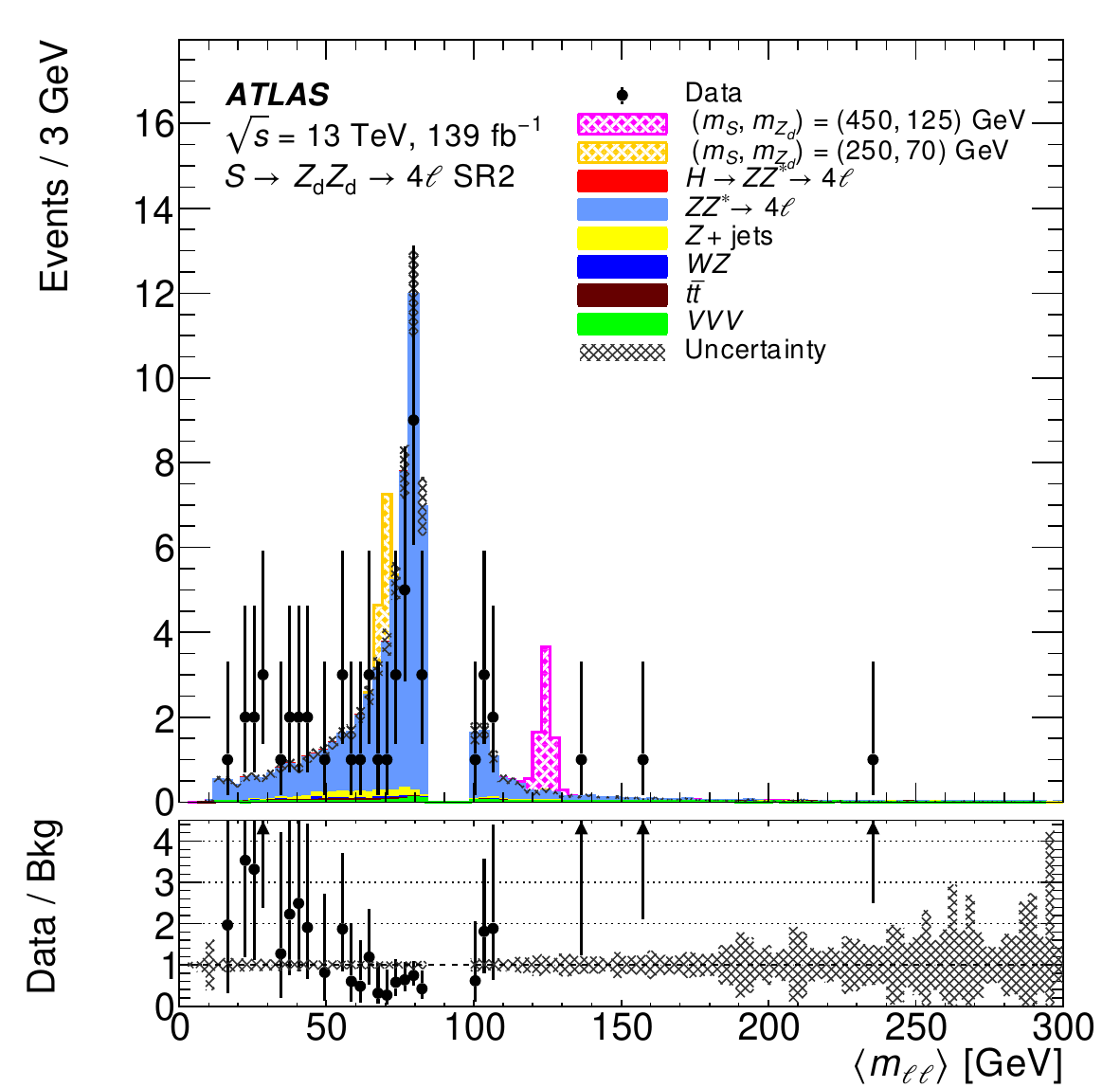}%
}
\caption{Distributions of the average dilepton mass
$\poleft\langle{}m_{\ell\ell}\poright\rangle = \half\poleft(m_{ab} + m_{cd}\poright)$
for the two signal regions (a)~SR1 and (b)~SR2.
The hatched bands show the uncertainty in the background prediction.
The uncertainties on the data points are asymmetric Poisson errors
(Eqs.~(40.82a) and~(40.82b) \ifnotpo{of Ref.~\cite{pdg24}}\ifpo{from the 2024 Review of Particle Properties}).
The lower panels show the ratio of the observed data to the
background (`Bkg') predictions.  The arrows at the upper edge indicate data points
that fall outside of the $y$-axis range, and the plotted uncertainties
include both the Poisson errors on the data and the uncertainty
on the background prediction.
Distributions for two example signal points are also shown.
These are normalized to production cross sections of
$\SI{0.5}{\femto\barn}$ for SR1 and
$\SI{0.1}{\femto\barn}$ for SR2.
\label{fig:averageDileptonMc16Full}}
\end{center}
\end{figure}

\begin{figure}[!tbp]
\begin{center}
\subfloat[]{\includegraphics[width=0.49\textwidth]{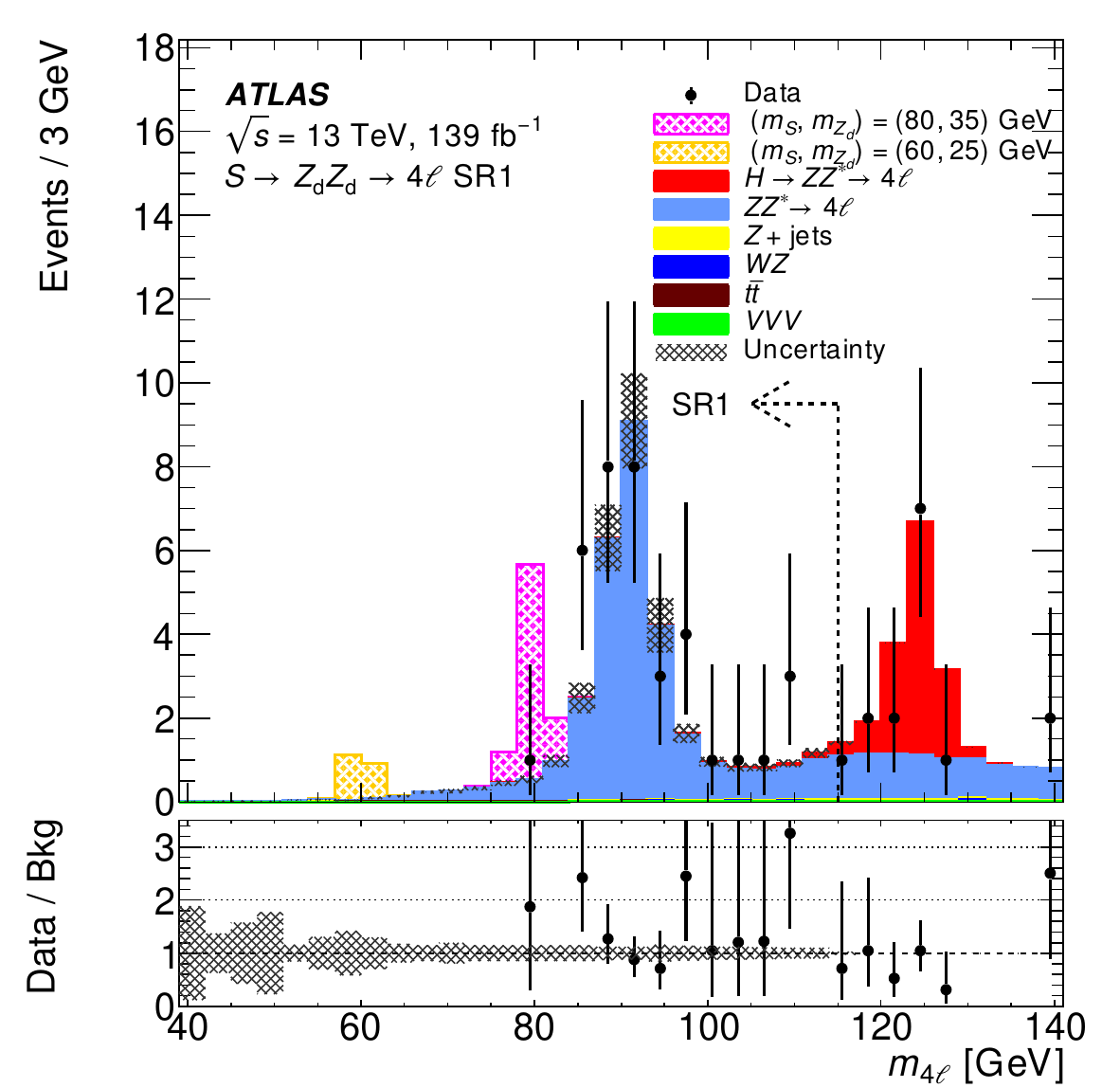}%
}
\subfloat[]{\includegraphics[width=0.49\textwidth]{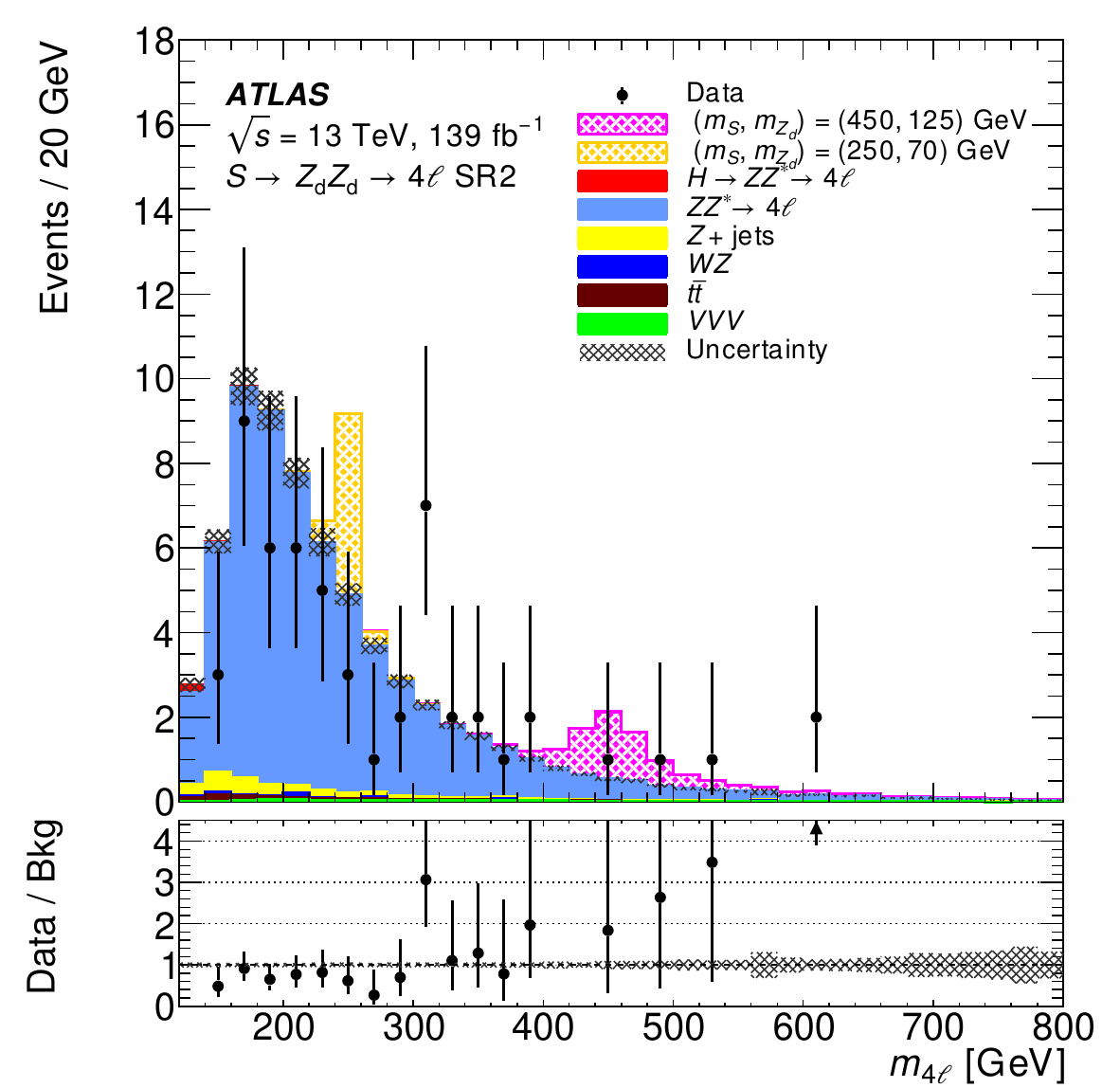}%
}
\caption{Distributions of the total invariant mass
$m_{4\ell}$
for the two signal regions (a)~SR1 and (b)~SR2.
In (a), the requirement $m_{4\ell} < \SI{115}{\GeV}$ is not applied,
but is shown by the arrow.
The hatched bands show the uncertainty in the background prediction;
however, in (a), background uncertainties
are not shown for $m_{4\ell} > \SI{115}{\GeV}$.
The uncertainties on the data points are asymmetric Poisson errors
(Eqs.~(40.82a) and~(40.82b) \ifnotpo{of Ref.~\cite{pdg24}}\ifpo{from the 2024 Review of Particle Properties}).
The lower panels show the ratio of the observed data to the
background (`Bkg') predictions.  The arrows at the upper edge indicate data points
that fall outside of the $y$-axis range, and the plotted uncertainties
include both the Poisson errors on the data and the uncertainty
on the background prediction.
Distributions for two example signal points are also shown.
These are normalized to production cross sections of
$\SI{0.5}{\femto\barn}$ for SR1 and
$\SI{0.1}{\femto\barn}$ for SR2.
\label{fig:M4lMc16Full}}
\end{center}
\end{figure}

\begin{figure}[!tbp]
\begin{center}
\subfloat[]{\includegraphics[width=0.49\textwidth]{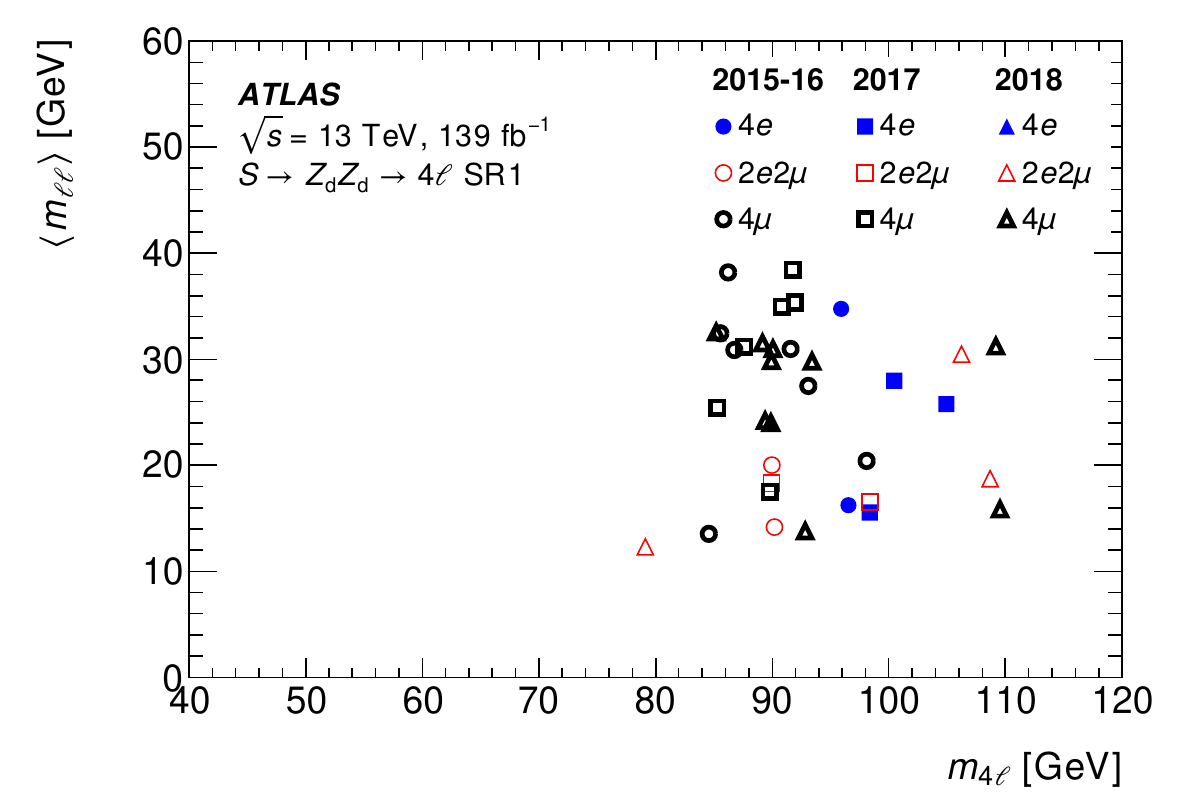}%
}
\subfloat[]{\includegraphics[width=0.49\textwidth]{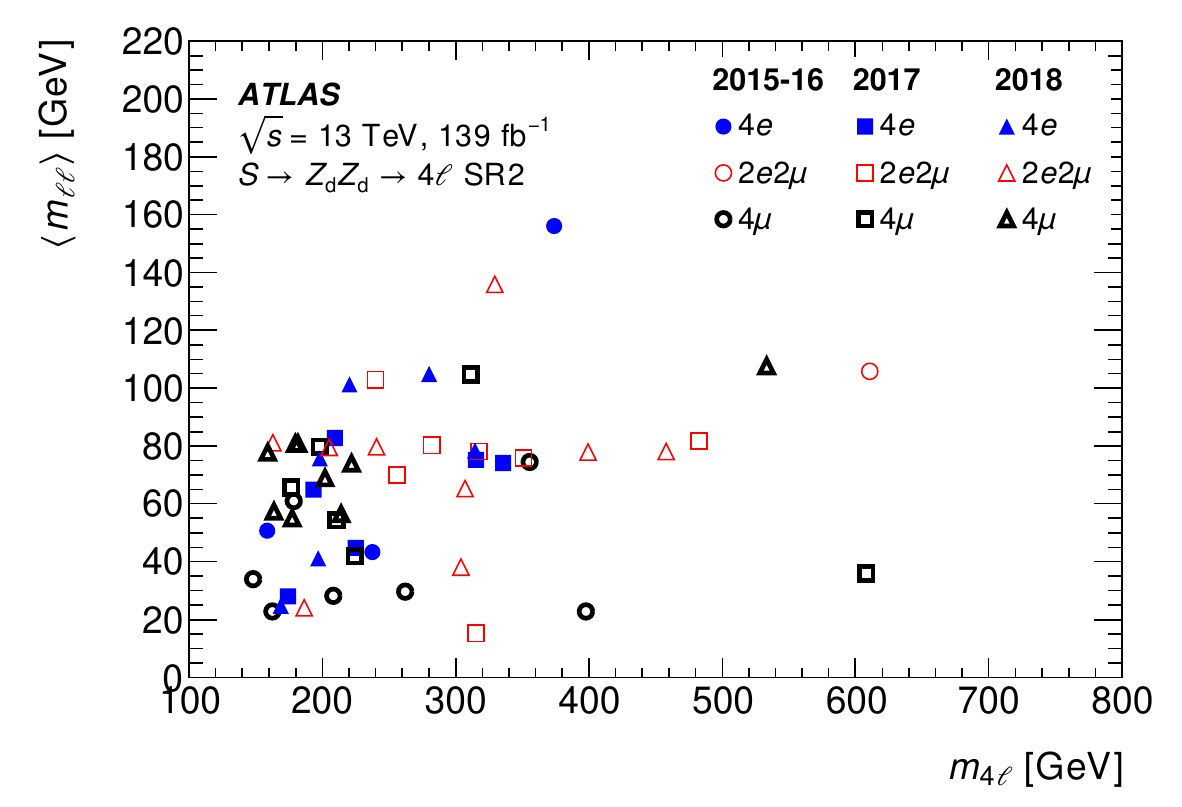}%
}
\caption{Events selected in signal regions (a)~SR1 and (b)~SR2,
represented in the
$(m_{4\ell}, \langle{}m_{\ell\ell}\rangle)$ plane.
The points are differentiated by final state ($4e$, $2e2\mu$, and $4\mu$)
and by data-taking period (2015--2016, 2017, and 2018).
\label{fig:MllM4lMc16Full}}
\end{center}
\end{figure}

\begin{figure}[!tbp]
\begin{center}
\subfloat[]{\includegraphics[width=0.5\textwidth]{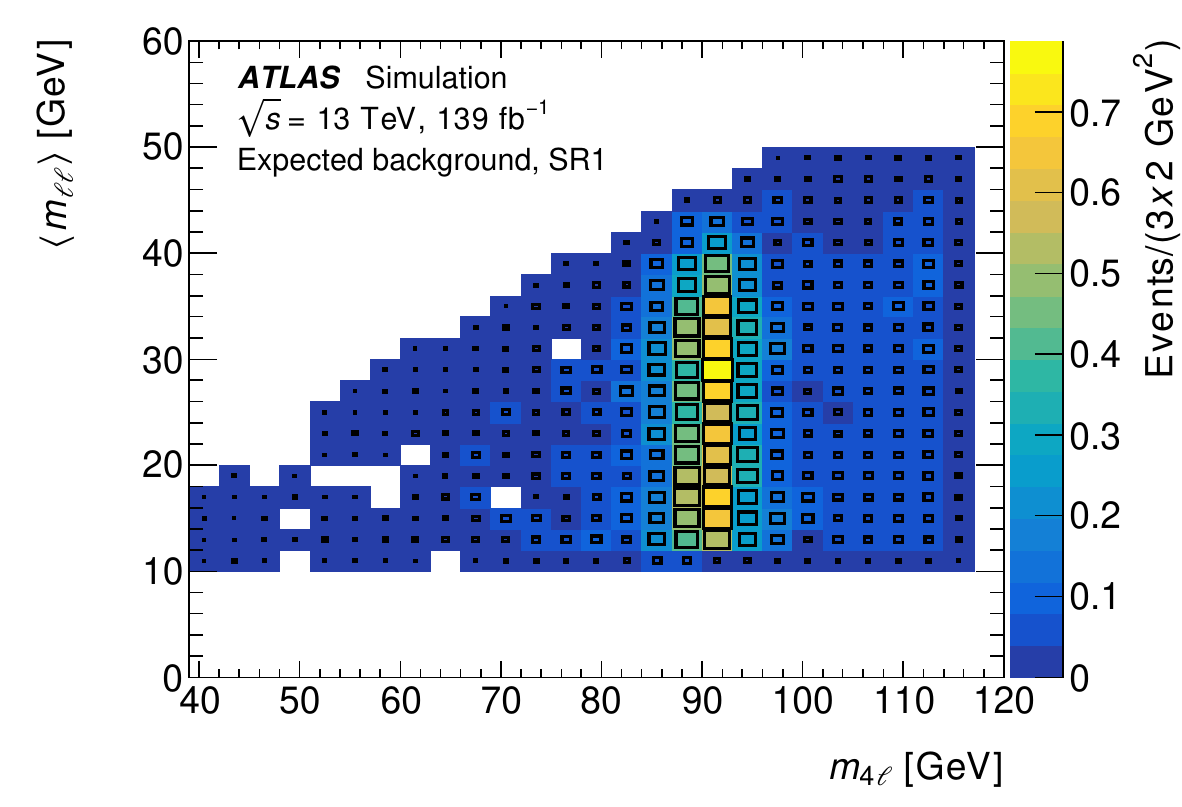}%
}
\subfloat[]{\includegraphics[width=0.5\textwidth]{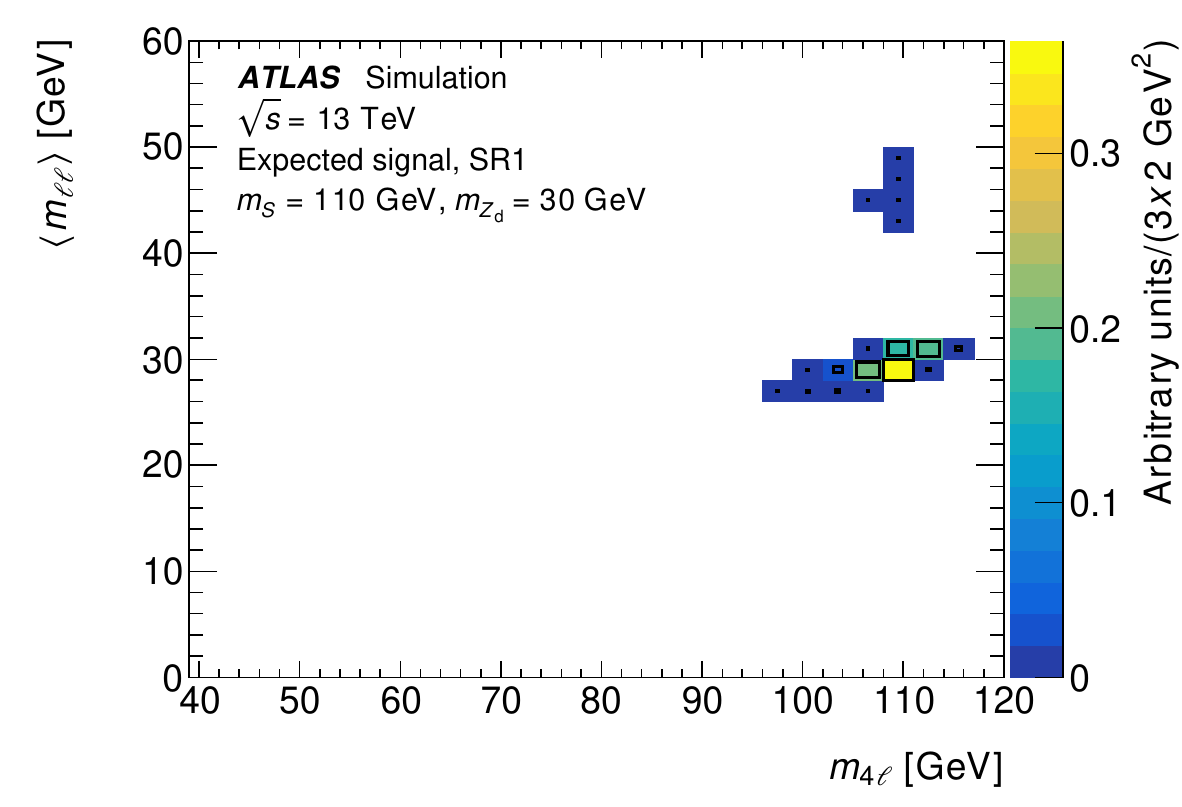}%
}\\
\subfloat[]{\includegraphics[width=0.5\textwidth]{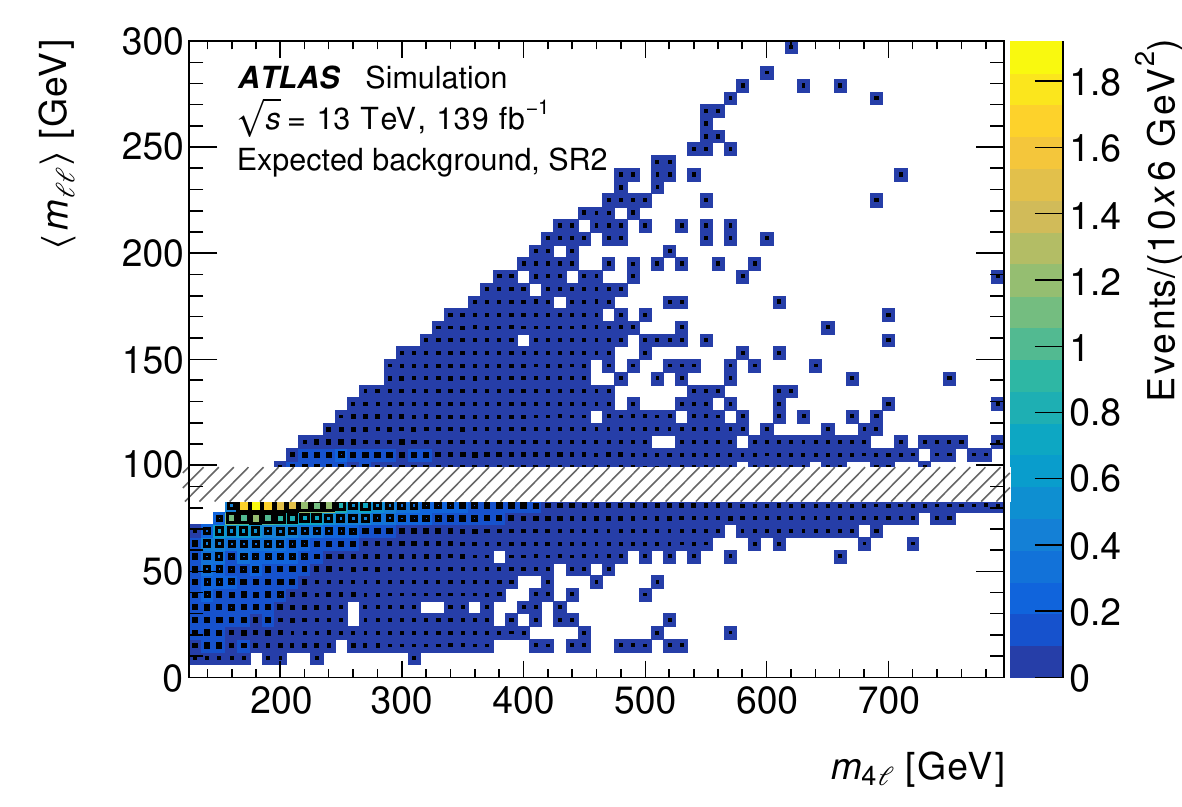}%
}
\subfloat[]{\includegraphics[width=0.5\textwidth]{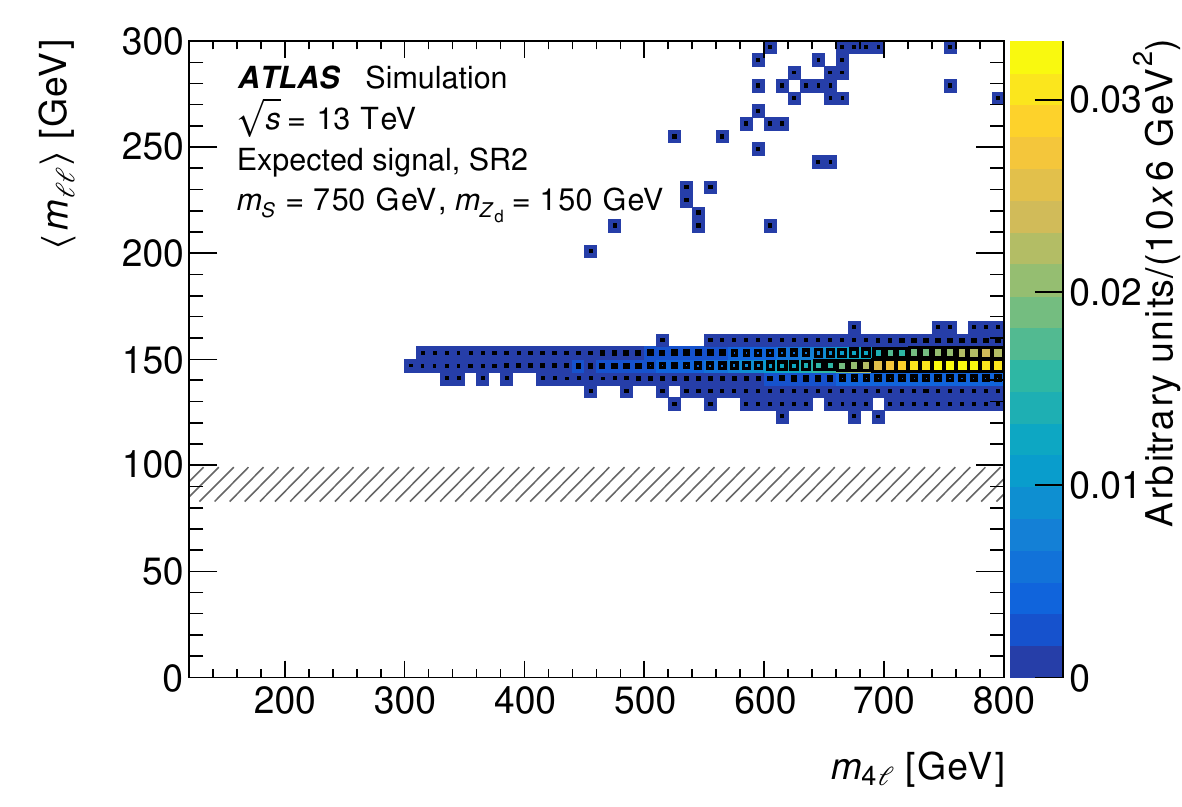}%
}
\caption{(a) Estimated background and (b) expected signal shape
for $(m_{S}, m_{\Zd}) = (\SI{110}{\GeV},\ifpo{\ }\SI{30}{\GeV}$)
in SR1 in the
$(m_{4\ell}, \langle{}m_{\ell\ell}\rangle)$ plane,
using both histograms and box representations.
(c, d) The corresponding plots for SR2
with $(m_{S}, m_{\Zd}) = (\SI{750}{\GeV},\ifpo{\ }\SI{150}{\GeV}$).
The signal distributions are normalized to unit volume.
Bins with less than 0.1\% of the content of the largest bin
are drawn as white.
\label{fig:2dplots}}
\end{center}
\end{figure}

\cref{tab:bkgEstimatesAS} reports the observed and expected yields in the signal regions.
In SR1 ($m_{4\ell} < \SI{115}{\GeV}$), 36~events are observed with an estimated background
of $32.6\pm3.3$, while in SR2 ($m_{4\ell} > \SI{130}{\GeV}$), there are 55~events with
an estimated background of $66.8\pm8.0$.
The distributions of
$\langle m_{\ell\ell}\rangle = \frac{1}{2}\left(m_{ab} + m_{cd}\right)$ for
both signal regions are shown in \cref{fig:averageDileptonMc16Full},
while \cref{fig:M4lMc16Full} shows
the overall invariant mass $m_{4\ell}$, which can also be interpreted
as the mass of the scalar candidate.
The selected events are represented
in the $(m_{4\ell}, \langle m_{\ell\ell}\rangle)$ plane in \cref{fig:MllM4lMc16Full},
while the estimated backgrounds and example signal shapes in SR1 and SR2
are shown in \cref{fig:2dplots}.

\begin{table}[t]
\caption{Expected event yields of the SM background processes and observed data
in the two signal regions SR1 and SR2.
The first uncertainty is the statistical component of the total uncertainty
and the second the systematic component.}
\centering
\begin{tabular}{@{\extracolsep{\fill}} l
S@{\,}@{$\pm$}@{\,}S@{\,}@{$\pm$}@{\,}l
S@{\,}@{$\pm$}@{\,}S@{\,}@{$\pm$}@{\,}l}
\toprule
Process & \multicolumn{6}{c}{Yield ($\pm$\,stat.\,$\pm$\,syst.)}\bigstrut[t] \\
& \multicolumn{3}{c}{SR1} & \multicolumn{3}{c}{SR2}\bigstrut[b] \\
\hline
\bigstrut[t]
$ZZ^*\to 4\ell$      & 30.9 & 0.5  & 3.2      & 62.0 & 0.5  & 7.9  \\
$H\to ZZ^*\to 4\ell$ & 0.61 & 0.01 & 0.04     & 0.22 & 0.01 & 0.03 \\
$WZ$                 & 0.06 & 0.04 & 0.03     & 0.42 & 0.10 & 0.05 \\
VVV                  & 0.06 & 0.01 & 0.01     & 0.78 & 0.02 & 0.14 \\

$\ttbar$             & 0.33 &<0.01 & 0.31     & 0.78 & 0.01 & 0.46 \\
$Z+\textrm{jets}$    & 0.70 & 0.01 & 0.72     & 2.53 &<0.01 & 1.08 \\
\hline
Total\bigstrut       & 32.6 & 0.5  & 3.3      & 66.8 & 0.6  & 8.0  \\
\hline
Data\bigstrut        & \multicolumn{3}{l}{36} & \multicolumn{3}{l}{55}  \\
\bottomrule
\end{tabular}
\label{tab:bkgEstimatesAS}
\end{table}

A two-dimensional interpolation procedure is used to obtain the shape of the
$\langle m_{\ell\ell}\rangle$ distribution for the predicted signal
at any point on the $(m_{S}, m_{\Zd})$ plane.
For each generated
signal MC sample, the $\langle m_{\ell\ell}\rangle$ distribution is fit.
In SR1, these distributions are fit well by Gaussians,
with three parameters.  However, in SR2, the distributions are observed
to be asymmetric, and a double-sided Crystal Ball distribution~\cite{Oreglia:1980cs} is used
instead, with
seven parameters.  For this interpolation, the
$S$~boson production cross-section and decay branching ratios are
factored out, with the cross-section taken from Ref.~\cite{deFlorian:2016spz}
and the branching ratio set to one.  Each fit parameter is then interpolated
separately using the thin plate spline
method~\cite{tps-code,tps-duchon,tps-bookstein},
with the regularization parameter set to zero so that the interpolation
passes smoothly through each control point.
The interpolation procedure is validated by removing points from the
interpolation one by one and then comparing the $\langle m_{\ell\ell}\rangle$
distribution for that point to what would be interpolated.

\begin{figure}[!tbp]
\begin{center}
\subfloat[]{\includegraphics[width=0.33\textwidth]{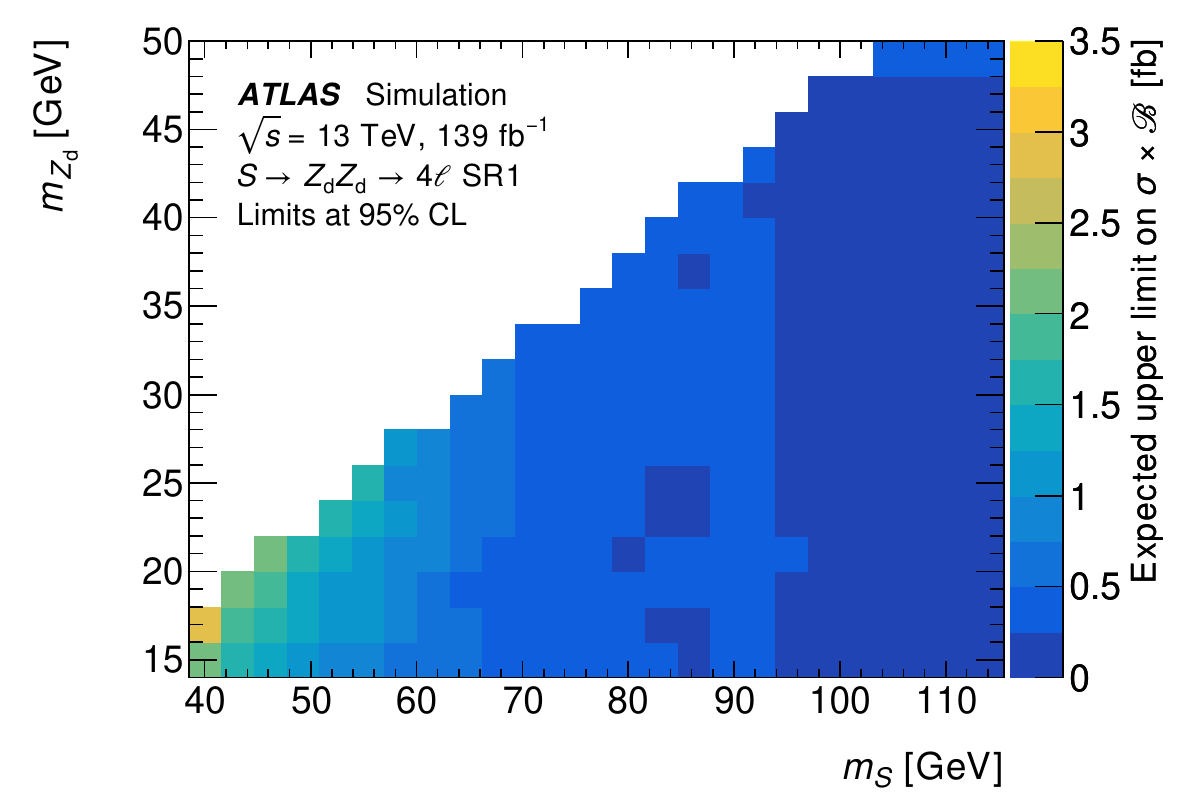}%
\label{fig:ASSR1_expected_Limit}}
\subfloat[]{\includegraphics[width=0.33\textwidth]{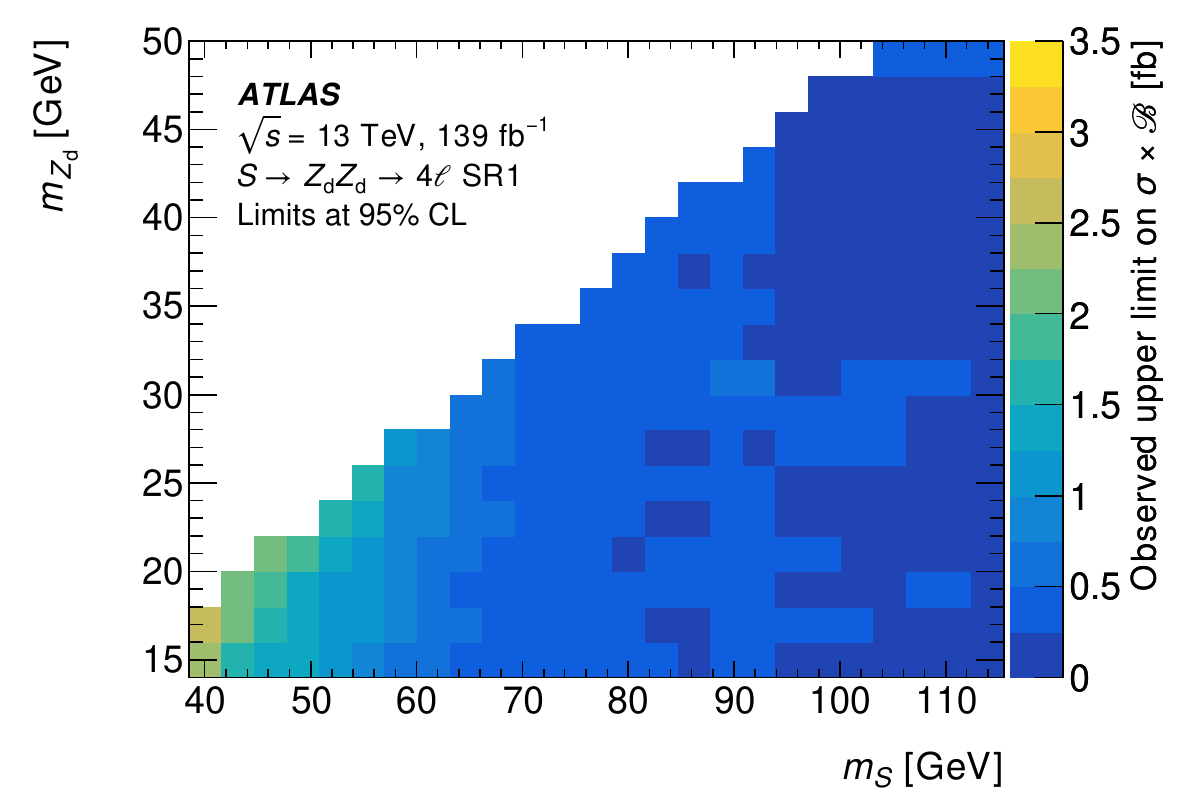}%
\label{fig:ASSR1_observed_Limit}}
\subfloat[]{\includegraphics[width=0.33\textwidth]{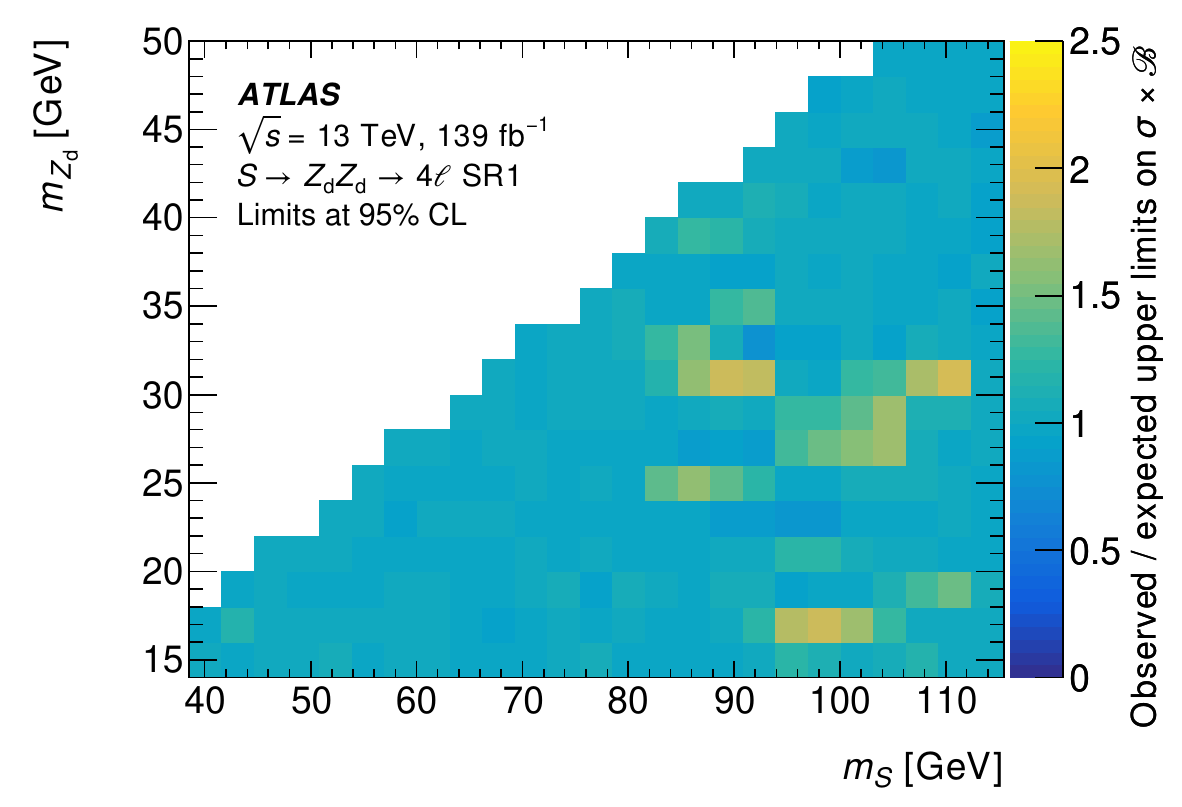}%
\label{fig:ASSR1_Limit_Ratio}} \\

\subfloat[]{\includegraphics[width=0.33\textwidth]{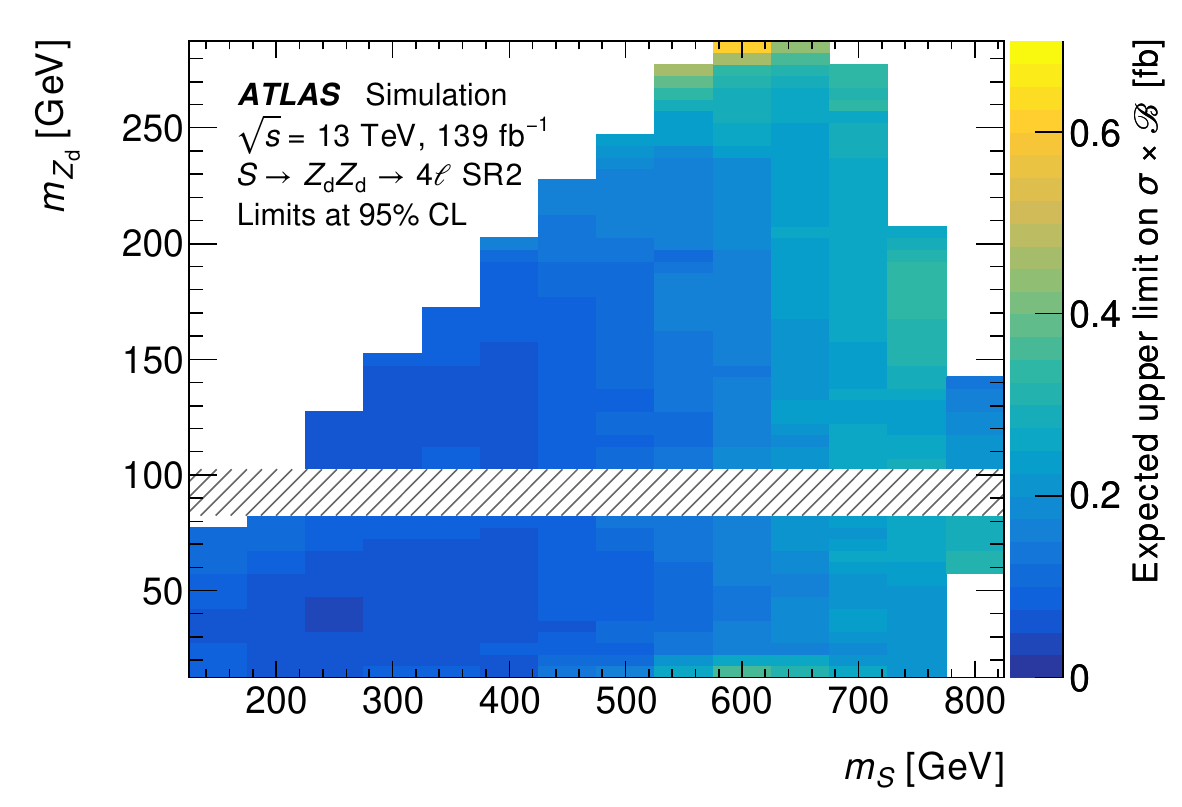}%
\label{fig:ASSR2_expected_Limit}}
\subfloat[]{\includegraphics[width=0.33\textwidth]{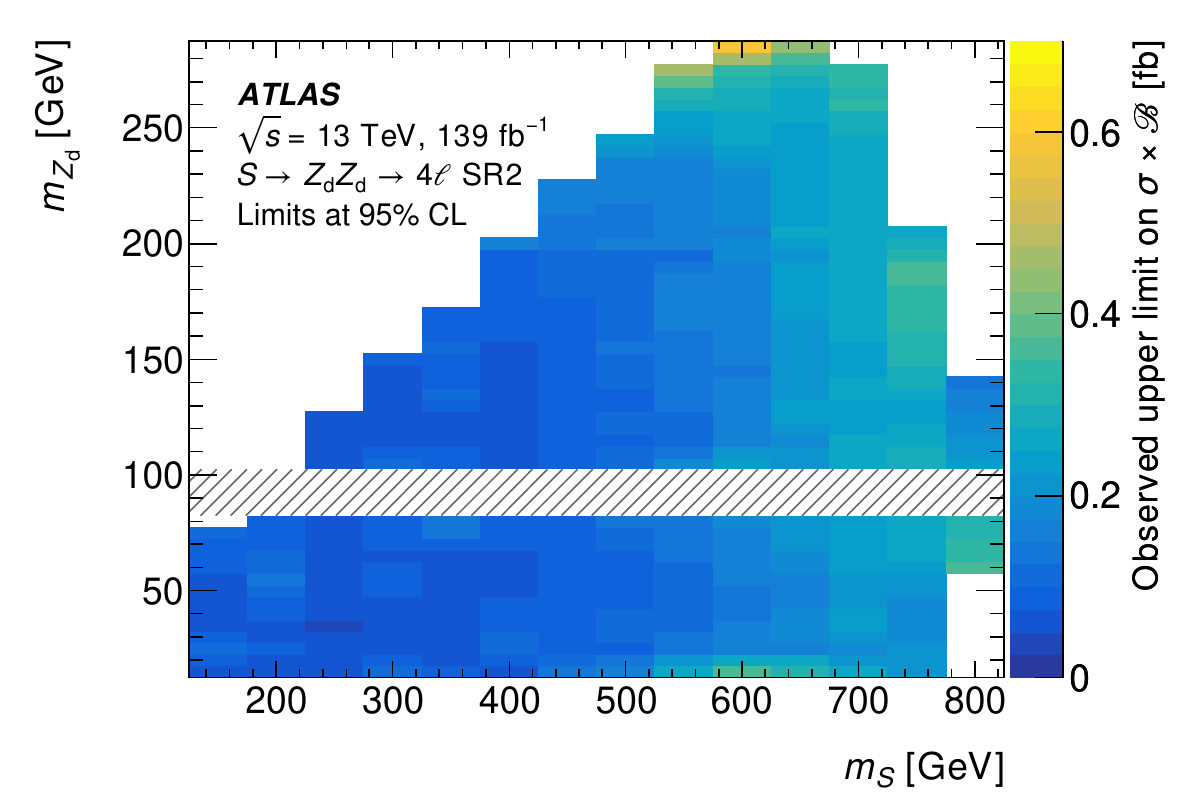}%
\label{fig:ASSR2_observed_Limit}}
\subfloat[]{\includegraphics[width=0.33\textwidth]{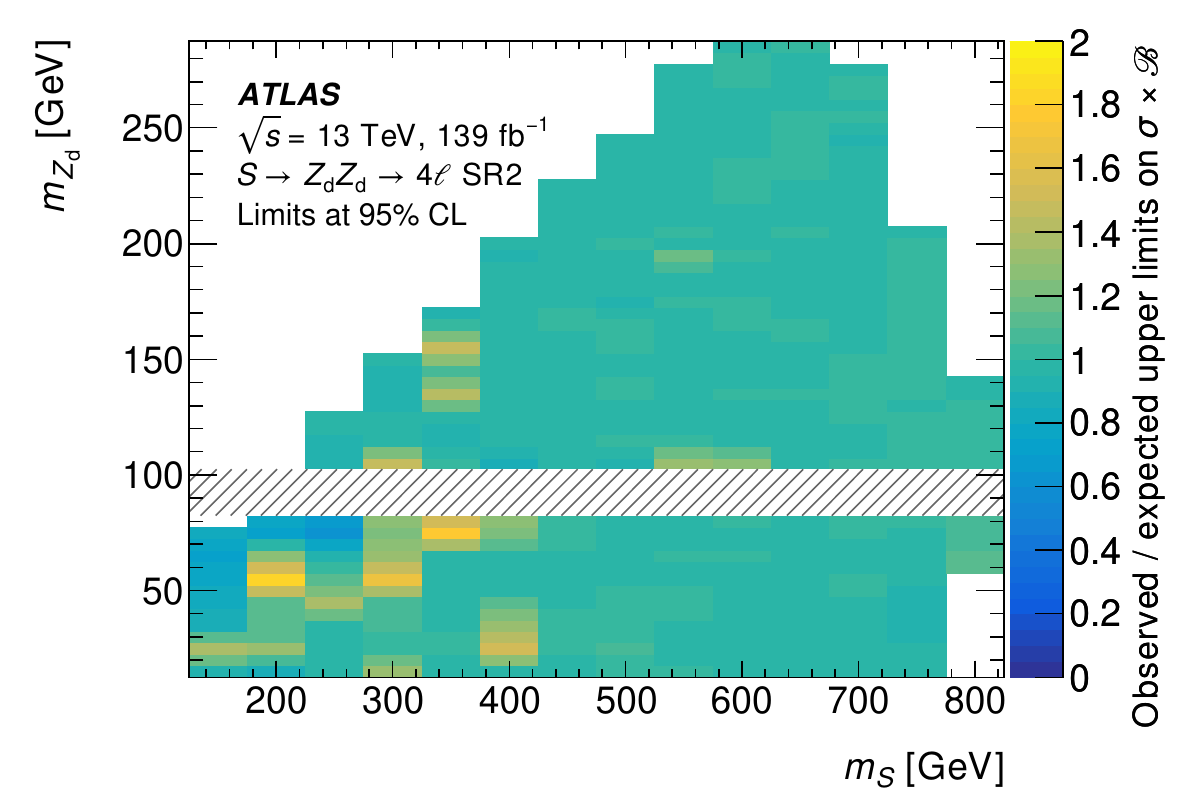}%
\label{fig:ASSR2_Limit_Ratio}} \\
\caption{95\popercent CL expected and observed limits on the
total cross-section times branching ratio
$\sigma(gg \to S) \times \mathcal{B}(S\to \Zd\Zd \to 4\ell)$.
(a)~SR1~expected limit; (b)~SR1~observed limit;
(c)~ratio of observed to expected limits in SR1;
(d)~SR2~expected limit; (e)~SR2~observed limit;
(f)~ratio of observed to expected limits in SR2.
The horizontal bands in the SR2 figures show the region
excluded by the $Z$ boson veto requirement,
$|m_{ab,cd} - m_Z| > \SI{8}{\GeV}$.
}
\end{center}
\end{figure}

\begin{figure}[!tbp]
\begin{center}
\subfloat[]{\includegraphics[width=0.49\textwidth]{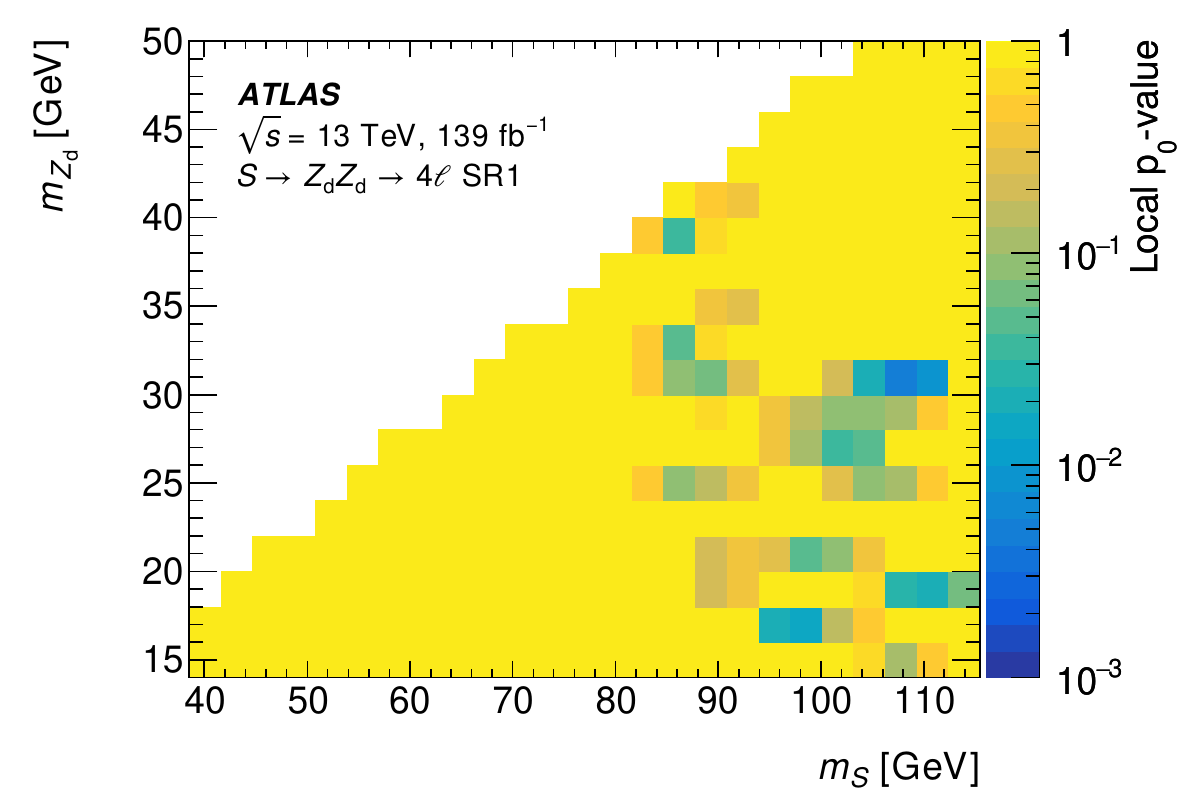}\label{fig:2DPvalueSR2}
}
\subfloat[]{\includegraphics[width=0.49\textwidth]{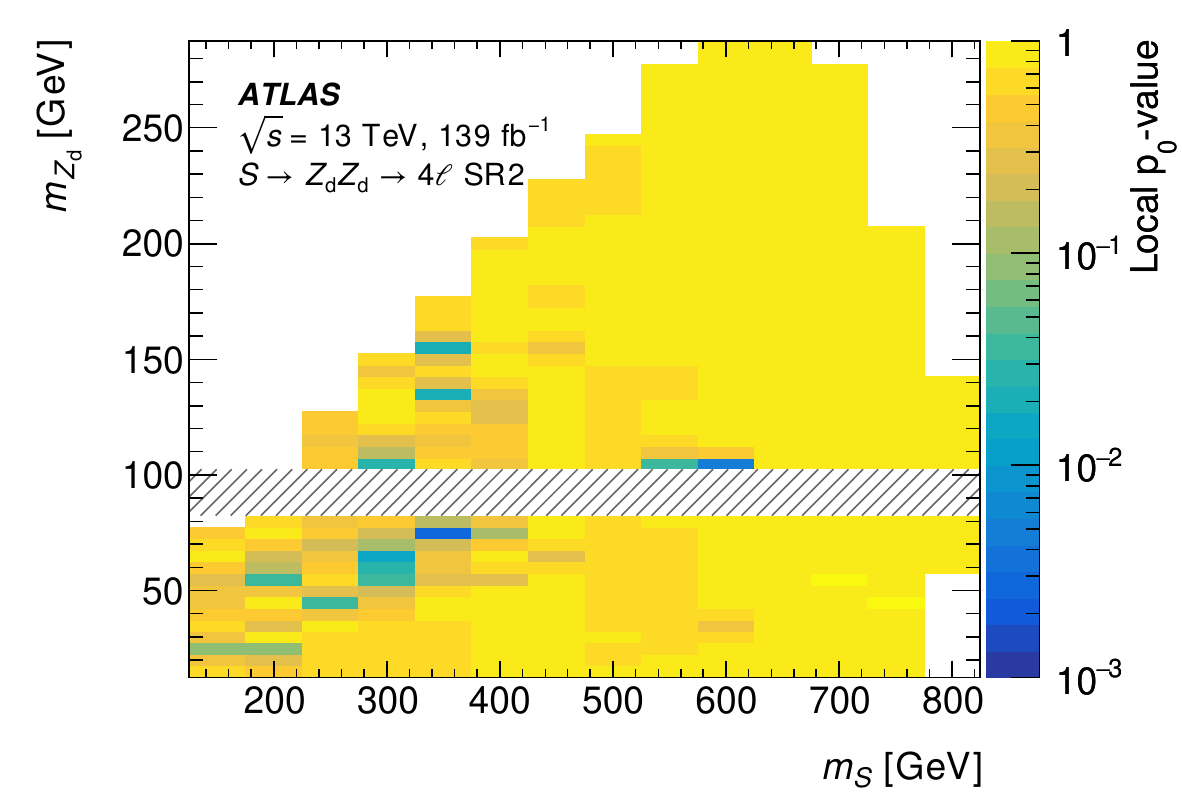}\label{fig:2DPvalueSR1}
} \\
\caption{Local $p_0$-values in the $(m_{4\ell}, \langle{}m_{\ell\ell}\rangle)$ plane for
(a) SR1 and (b) SR2, evaluated in slices in $m_{4\ell}$.
The horizontal band in the SR2 figure shows the region
excluded by the $Z$ boson veto requirement,
$|m_{ab,cd} - m_Z| > \SI{8}{\GeV}$.
}
\end{center}
\end{figure}

Exclusion limits on %
$\sigma(gg \to S) \times \mathcal{B}(S\rightarrow \Zd\Zd \rightarrow 4\ell)$
are set using frequentist significance testing of a
statistic derived from a profile likelihood ratio that considers
the data, the background model, and the signal model distributions
for various hypothesized $(m_{S}, m_{\Zd})$ points~\cite{Cowan:2010js-witherratum, Read:2002hq, cranmer2012histfactory}.
To reduce the resources required, events are
binned coarsely in $m_{4\ell}$ and limits set independently within each
$m_{4\ell}$ bin as a function of~$\langle m_{\ell\ell}\rangle$.
Bins with a width of $\SI{1}{\GeV}$ are used for $\langle m_{\ell\ell}\rangle$.
In SR1 there are eight~equal-sized bins in $m_{4\ell}$ in the range of
$\SI{30}{\GeV}$ to $\SI{110}{\GeV}$, and for SR2, there are
14~equal-sized bins in the range of $\SI{125}{\GeV}$ to $\SI{825}{\GeV}$.
These bin widths were chosen to have sufficient statistics while still being
commensurate with the widths of the new scalar and the dark vector boson.
Only signal MC events with
$m_{4\ell}$ within one bin width of the generated $m_S$ value
are used to form the signal model distributions for this calculation.
Limits are evaluated as described
in Refs.~\cite{HDBS-2018-55,EXOT-2016-22,EXOT-2013-15}
(using the \verb|HypoTestInverter| component of
the \ROOSTATS toolkit~\cite{Moneta:2010pm} and evaluating the test statistic
distributions with MC pseudoexperiments
taking into account both statistical and systematic uncertainties),
resulting in CL$_\text{s}$ frequentist upper limits at
95\% confidence level (CL) on
$\sigma(gg \to S) \times \mathcal{B}(S\rightarrow \Zd\Zd \rightarrow 4\ell)$
as a function of the $m_{\Zd}$ parameter in each $m_{4\ell}$ bin.
Expected and observed limits and their ratios are shown in
\cref{fig:ASSR1_expected_Limit,fig:ASSR1_observed_Limit,fig:ASSR1_Limit_Ratio} for SR1 and
\cref{fig:ASSR2_expected_Limit,fig:ASSR2_observed_Limit,fig:ASSR2_Limit_Ratio} for SR2.
The corresponding one-dimensional local $p_0$-values
are shown in \cref{fig:2DPvalueSR1,fig:2DPvalueSR2}.
In SR1, the point with the smallest $p_0$-value is at
$(m_{S}, m_{\Zd}) = (\SI{110}{\GeV},\SI{30}{\GeV})$ with a local
significance of $2.7\sigma$, while in SR2 the smallest
$p_0$-value is at $(m_{S}, m_{\Zd}) = (\SI{350}{\GeV},\SI{75}{\GeV})$
with a local significance of $2.8\sigma$.
The local significance can be converted to a global significance by
including the look-elsewhere effect using
MC pseudo experiments as outlined in Ref.~\cite{Vitells_2011}.
For SR2, the largest global significance is $0.5\sigma$. %

The local significance of the deviation in SR1 is re-evaluated
using the same frequentist significance testing as described above but
applied in the two-dimensional $(m_{4\ell},\langle m_{\ell\ell}\rangle)$ plane
with a binning of $\Delta m_{4\ell}\times \Delta \langle m_{\ell\ell}\rangle = \SI{3}{\GeV} \times \SI{1}{\GeV}$.  The evaluation is done in a narrow window
scanned over the plane with size set to contain at least~99\% of the
expected signal.
The resulting local $p_0$-value still corresponds to a significance of~$2.7\sigma$ but for the mass point
$(m_{S}, m_{\Zd}) = (\SI{109}{\GeV},\SI{31}{\GeV})$.
Converting this to a global significance yields~$1.6\sigma$.

%
%
%
%
%
%
%
%
%
%
%
%
%
%
%
%
%
%
%
%
%
%
%
%
%
%
%
%
%
%
%
%
%
%
%
%
%
%
%
%
%
%
%
%
%
%

%
%
%
%
%
%

%
%
%
%
%
%
%

%
%
%
%
%
%
%
%
%
%
%
%
%

%
%
%
%
%

%
%
%
%
%
%
%
%
%
%
%
%
%
%
%
%
%
%
%
%
%
%
%
%
%
%
%
%
%
%
%
%
%


%
%
\FloatBarrier

\section{Conclusions}

A search is presented
for the decay of a new scalar~$S$
into two new spin-1 particles $S\rightarrow \Zd\Zd$, with each
$\Zd$~boson decaying promptly into a pair of electrons or muons, yielding an inclusive
four-lepton final state.  It is conducted
in the plane of the reconstructed four-lepton mass and the average dilepton mass of the $\Zd \rightarrow \ell\ell$ candidates,
and it uses \SI{139}{\ifb} of
$pp$ collision data at $\sqrt{s}=\SI{13}{\TeV}$
recorded by the ATLAS experiment at the
LHC during the period 2015--2018.  Two different signal regions are studied,
corresponding to $\SI{30}{\GeV} < m_S < \SI{115}{\GeV}$ and
$\SI{130}{\GeV} < m_S < \SI{800}{\GeV}$.
The data are found to be consistent with the Standard Model background expectation, and 95\% CL
upper limits are set on the total cross-section times branching ratio
$\sigma(gg \to S) \times \mathcal{B}(S\rightarrow \Zd\Zd \rightarrow 4\ell)$ as
a function of $m_S$ and~$m_{\Zd}$.  In the first signal region,
the limits range from
$\SI{0.14}{\fb}$ to $\SI{3.1}{\fb}$ and in the second, from
$\SI{0.05}{\fb}$ to $\SI{0.60}{\fb}$.
These constraints apply to the dark sector described by the Hidden Abelian
Higgs Model, as well as to similar models resulting in a four-lepton final
state.

%
%
%
%


%
\section*{Acknowledgements}
%

%
%

%
%

We thank CERN for the very successful operation of the LHC and its injectors, as well as the support staff at
CERN and at our institutions worldwide without whom ATLAS could not be operated efficiently.

The crucial computing support from all WLCG partners is acknowledged gratefully, in particular from CERN, the ATLAS Tier-1 facilities at TRIUMF/SFU (Canada), NDGF (Denmark, Norway, Sweden), CC-IN2P3 (France), KIT/GridKA (Germany), INFN-CNAF (Italy), NL-T1 (Netherlands), PIC (Spain), RAL (UK) and BNL (USA), the Tier-2 facilities worldwide and large non-WLCG resource providers. Major contributors of computing resources are listed in Ref.~\cite{ATL-SOFT-PUB-2025-001}.

We gratefully acknowledge the support of ANPCyT, Argentina; YerPhI, Armenia; ARC, Australia; BMWFW and FWF, Austria; ANAS, Azerbaijan; CNPq and FAPESP, Brazil; NSERC, NRC and CFI, Canada; CERN; ANID, Chile; CAS, MOST and NSFC, China; Minciencias, Colombia; MEYS CR, Czech Republic; DNRF and DNSRC, Denmark; IN2P3-CNRS and CEA-DRF/IRFU, France; SRNSFG, Georgia; BMBF, HGF and MPG, Germany; GSRI, Greece; RGC and Hong Kong SAR, China; ISF and Benoziyo Center, Israel; INFN, Italy; MEXT and JSPS, Japan; CNRST, Morocco; NWO, Netherlands; RCN, Norway; MEiN, Poland; FCT, Portugal; MNE/IFA, Romania; MESTD, Serbia; MSSR, Slovakia; ARIS and MVZI, Slovenia; DSI/NRF, South Africa; MICINN, Spain; SRC and Wallenberg Foundation, Sweden; SERI, SNSF and Cantons of Bern and Geneva, Switzerland; NSTC, Taipei; TENMAK, T\"urkiye; STFC/UKRI, United Kingdom; DOE and NSF, United States of America.

Individual groups and members have received support from BCKDF, CANARIE, CRC and DRAC, Canada; CERN-CZ, PRIMUS 21/SCI/017 and UNCE SCI/013, Czech Republic; COST, ERC, ERDF, Horizon 2020, ICSC-NextGenerationEU and Marie Sk{\l}odowska-Curie Actions, European Union; Investissements d'Avenir Labex, Investissements d'Avenir Idex and ANR, France; DFG and AvH Foundation, Germany; Herakleitos, Thales and Aristeia programmes co-financed by EU-ESF and the Greek NSRF, Greece; BSF-NSF and MINERVA, Israel; Norwegian Financial Mechanism 2014-2021, Norway; NCN and NAWA, Poland; La Caixa Banking Foundation, CERCA Programme Generalitat de Catalunya and PROMETEO and GenT Programmes Generalitat Valenciana, Spain; G\"{o}ran Gustafssons Stiftelse, Sweden; The Royal Society and Leverhulme Trust, United Kingdom.

In addition, individual members wish to acknowledge support from Chile: Agencia Nacional de Investigaci\'on y Desarrollo (FONDECYT 1190886, FONDECYT 1210400); China: National Natural Science Foundation of China (NSFC - 12175119, NSFC 12275265, NSFC-12075060); Czech Republic: PRIMUS Research Programme (PRIMUS/21/SCI/017); EU: H2020 European Research Council (ERC - 101002463, H2020-MSCA-IF-2020: HPOFHIC - 10103); European Union: European Research Council (ERC - 948254), Horizon 2020 Framework Programme (MUCCA - CHIST-ERA-19-XAI-00), European Union, Future Artificial Intelligence Research (FAIR-NextGenerationEU PE00000013), Italian Center for High Performance Computing, Big Data and Quantum Computing (ICSC, NextGenerationEU), Marie Sklodowska-Curie Actions (EU H2020 MSC IF GRANT NO 101033496); France: Agence Nationale de la Recherche (ANR-20-CE31-0013, ANR-21-CE31-0013, ANR-21-CE31-0022), Investissements d'Avenir Idex (ANR-11-LABX-0012), Investissements d'Avenir Labex (ANR-11-LABX-0012); Germany: Baden-Württemberg Stiftung (BW Stiftung-Postdoc Eliteprogramme), Deutsche Forschungsgemeinschaft (DFG - CR 312/5-1); Italy: Istituto Nazionale di Fisica Nucleare (FELLINI G.A. n. 754496, ICSC, NextGenerationEU); Japan: Japan Society for the Promotion of Science (JSPS KAKENHI JP21H05085, JSPS KAKENHI JP22H01227, JSPS KAKENHI JP22H04944); Netherlands: Netherlands Organisation for Scientific Research (NWO Veni 2020 - VI.Veni.202.179); Norway: Research Council of Norway (RCN-314472); Poland: Polish National Agency for Academic Exchange (PPN/PPO/2020/1/00002/U/00001), Polish National Science Centre (NCN 2021/42/E/ST2/00350, NCN UMO-2019/34/E/ST2/00393, NCN \& H2020 MSCA 945339, UMO-2020/37/B/ST2/01043, UMO-2021/40/C/ST2/00187); Slovenia: Slovenian Research Agency (ARIS grant J1-3010); Spain: BBVA Foundation (LEO22-1-603), Generalitat Valenciana (Artemisa, FEDER, IDIFEDER/2018/048), La Caixa Banking Foundation (LCF/BQ/PI20/11760025), Ministry of Science and Innovation (MCIN \& NextGenEU PCI2022-135018-2, MICIN \& FEDER PID2021-125273NB, RYC2019-028510-I, RYC2020-030254-I, RYC2021-031273-I, RYC2022-038164-I), PROMETEO and GenT Programmes Generalitat Valenciana (CIDEGENT/2019/023, CIDEGENT/2019/027); Sweden: Swedish Research Council (VR 2018-00482, VR 2021-03651, VR 2022-03845, VR 2022-04683), Knut and Alice Wallenberg Foundation (KAW 2017.0100, KAW 2018.0157, KAW 2018.0458, KAW 2019.0447); Switzerland: Swiss National Science Foundation (SNSF - PCEFP2\_194658); United Kingdom: Leverhulme Trust (Leverhulme Trust RPG-2020-004); United States of America: U.S. Department of Energy (ECA DE-AC02-76SF00515), Neubauer Family Foundation.

%
%


%
\clearpage
\appendix
\printbibliography

@article{LHCB-PAPER-2017-038,
    xauthor = "Aaij, Roel and others",
    author = "{LHCb Collaboration}",
    title = "{Search for Dark Photons Produced in 13 TeV $pp$ Collisions}",
    eprint = "1710.02867",
    archivePrefix = "arXiv",
    primaryClass = "hep-ex",
    reportNumber = "LHCB-PAPER-2017-038, CERN-EP-2017-248",
    doi = "10.1103/PhysRevLett.120.061801",
    journal = "\PRL",
    volume = "120",
    number = "6",
    pages = "061801",
    year = "2018"
}

@article{Curtin:2014cca,
      author         = "Curtin, David and Essig, Rouven and Gori, Stefania and
                        Shelton, Jessie",
      title          = "{Illuminating dark photons with high-energy colliders}",
      journal        = "JHEP",
      volume         = "02",
      year           = "2015",
      pages          = "157",
      doi            = "10.1007/JHEP02(2015)157",
      eprint         = "1412.0018",
      archivePrefix  = "arXiv",
      primaryClass   = "hep-ph",
      reportNumber   = "YITP-SB-14-49",
      SLACcitation   = "%%CITATION = ARXIV:1412.0018;%%"
}

@article{Davoudiasl:2012ag,
      author         = "Davoudiasl, Hooman and Lee, Hye-Sung and Marciano,
                        William J.",
      title          = "{``Dark'' $Z$ implications for parity violation, rare meson
                        decays, and Higgs physics}",
      journal        = "{\PRD{}}",
      volume         = "85",
      year           = "2012",
      pages          = "115019",
      doi            = "10.1103/PhysRevD.85.115019",
      eprint         = "1203.2947",
      archivePrefix  = "arXiv",
      primaryClass   = "hep-ph",
      SLACcitation   = "%%CITATION = ARXIV:1203.2947;%%"
}

@article{pdg24,
    author = "{Particle Data Group} and Navas, S. and others",
    title = "{Review of particle physics}",
    doi = "10.1103/PhysRevD.110.030001",
    journal = "Phys. Rev. D",
    volume = "110",
    number = "3",
    pages = "030001",
    year = "2024"
}

@article{Vitells_2011,
	author="Vitells, Ofer and Gross, Eilam",
        title="{Estimating the significance of a signal in a multi-dimensional search}",
	journal="Astropart. Phys.",
	volume="35",
	year="2011",
	pages="230",
	doi="10.1016/j.astropartphys.2011.08.005",
        eprint         = "1105.4355",
        archivePrefix  = "arXiv",
        primaryClass   = "astro-ph"
}

@article{tps-bookstein,
  author={Bookstein, Fred L.},
  journal={IEEE Trans. Pattern Anal. Machine Intell.}, 
  title={Principal warps: thin-plate splines and the decomposition of deformations}, 
  year={1989},
  volume={11},
  number={6},
  pages={567-585},
  doi={10.1109/34.24792}}

@article{tps-duchon,
  title={Interpolation des fonctions de deux variables suivant le principe de la flexion des plaques minces},
  author={Duchon, Jean},
  journal={RAIRO Analyse num{\'e}rique},
  volume={10},
  number={R3},
  pages={5--12},
  year={1976},
  url={http://www.numdam.org/item?id=M2AN_1976__10_3_5_0},
}

@misc{tps-code,
      author = "Elonen, Jarno",
      title = "Thin Plate Spline editor --- an example program in C++",
      url = "https://elonen.iki.fi/code/tpsdemo"
      }

@misc{hahm,
      author = "Curtin, David and others",
      title = "SM + Dark Vector + Dark Higgs Madgraph5 Model",
      url="https://feynrules.irmp.ucl.ac.be/wiki/HAHM",
      accessed = "2023-12-01"
      }

@article{Hoeche:2009xc,
    author = "H{\"o}che, Stefan and Schumann, Steffen and Siegert, Frank",
    title = "{Hard photon production and matrix-element parton-shower merging}",
    eprint = "0912.3501",
    archivePrefix = "arXiv",
    primaryClass = "hep-ph",
    reportNumber = "ZU-TH-19-09, IPPP-09-94, DCPT-09-188, HD-THEP-09-28, MCNET-09-18",
    doi = "10.1103/PhysRevD.81.034026",
    journal = "Phys. Rev. D",
    volume = "81",
    pages = "034026",
    year = "2010"
}

@phdthesis{Oreglia:1980cs, 
  author         = "Oreglia, M.", 
  title          = "{A Study of the Reactions $\psi^\prime \to \gamma \gamma \psi$}", 
  school         = "SLAC", 
  url            = "http://www.slac.stanford.edu/pubs/slacreports/slac-r-236.html", 
  year           = "1980", 
  reportNumber   = "SLAC-R-236", 
}

@Article{Butterworth:2015oua,
      author         = "Butterworth, Jon and others",
      title          = "{PDF4LHC recommendations for LHC Run II}",
      journal        = "J. Phys. G",
      volume         = "43",
      year           = "2016",
      pages          = "023001",
      doi            = "10.1088/0954-3899/43/2/023001",
      eprint         = "1510.03865",
      archivePrefix  = "arXiv",
      primaryClass   = "hep-ph",
      reportNumber   = "OUTP-15-17P, SMU-HEP-15-12, TIF-UNIMI-2015-14,
                        LCTS-2015-27, CERN-PH-TH-2015-249",
      SLACcitation   = "%%CITATION = ARXIV:1510.03865;%%"
}

@article{Ball:2012cx,
      author         = "{NNPDF Collaboration} and Ball, Richard D. and others",
      title          = "{Parton distributions with LHC data}",
      journal        = "Nucl. Phys. B",
      volume         = "867",
      year           = "2013",
      pages          = "244",
      doi            = "10.1016/j.nuclphysb.2012.10.003",
      eprint         = "1207.1303",
      archivePrefix  = "arXiv",
      primaryClass   = "hep-ph",
      reportNumber   = "EDINBURGH-2012-08, IFUM-FT-997, FR-PHENO-2012-014,
                        RWTH-TTK-12-25, CERN-PH-TH-2012-037, SFB-CPP-12-47\,
                        --CERN-PH-TH-2012-037",
      SLACcitation   = "%%CITATION = ARXIV:1207.1303;%%"
}

@Article{Sjostrand:2007gs,
	author    = "Sj{\"o}strand, T. and Mrenna, S. and Skands, P.",
	title     = "{A brief introduction to PYTHIA 8.1}",
	journal   = "Comput. Phys. Commun.",
	volume    = "178",
	year      = "2008",
	pages     = "852-867",
	eprint    = "0710.3820",
	archivePrefix = "arXiv",
	primaryClass  =  "hep-ph",
	doi       = "10.1016/j.cpc.2008.01.036",
	SLACcitation  = "%%CITATION = 0710.3820;%%"
}

@Article{Lange:2001uf,
      author         = "Lange, D. J.",
      title          = "{The EvtGen particle decay simulation package}",
      booktitle      = "{Proceedings, 7th International Conference on B physics
                        at hadron machines (BEAUTY 2000)}",
      journal        = "Nucl. Instrum. Meth. A",
      volume         = "462",
      year           = "2001",
      pages          = "152",
      doi            = "10.1016/S0168-9002(01)00089-4",
      SLACcitation   = "%%CITATION = NUIMA,A462,152;%%"
}

@Article{Alwall:2014hca,
      author         = "Alwall, J. and Frederix, R. and Frixione, S. and Hirschi,
                        V. and Maltoni, F. and Mattelaer, O. and Shao, H. -S. and
                        Stelzer, T. and Torrielli, P. and Zaro, M.",
      title          = "{The automated computation of tree-level and
                        next-to-leading order differential cross sections, and
                        their matching to parton shower simulations}",
      journal        = "JHEP",
      volume         = "07",
      year           = "2014",
      pages          = "079",
      doi            = "10.1007/JHEP07(2014)079",
      eprint         = "1405.0301",
      archivePrefix  = "arXiv",
      primaryClass   = "hep-ph",
      reportNumber   = "CERN-PH-TH-2014-064, CP3-14-18, LPN14-066, MCNET-14-09,
                        ZU-TH-14-14",
      SLACcitation   = "%%CITATION = ARXIV:1405.0301;%%"
}

@Article{Nason:2004rx,
      author         = "Nason, Paolo",
      title          = "{A new method for combining NLO QCD with shower Monte Carlo algorithms}",
      journal        = "JHEP",
      volume         = "11",
      pages          = "040",
      doi            = "10.1088/1126-6708/2004/11/040",
      year           = "2004",
      eprint         = "hep-ph/0409146",
      archivePrefix  = "arXiv",
}

@Article{Frixione:2007vw,
      author         = "Frixione, Stefano and Nason, Paolo and Oleari, Carlo",
      title          = "{Matching NLO QCD computations with parton shower
                        simulations: the POWHEG method}",
      journal        = "JHEP",
      volume         = "11",
      pages          = "070",
      doi            = "10.1088/1126-6708/2007/11/070",
      year           = "2007",
      eprint         = "0709.2092",
      archivePrefix  = "arXiv",
      primaryClass   = "hep-ph",
}

@Article{Alioli:2010xd,
      author         = "Alioli, Simone and Nason, Paolo and Oleari, Carlo and Re,
                        Emanuele",
      title          = "{A general framework for implementing NLO calculations in
                        shower Monte Carlo programs: the POWHEG BOX}",
      journal        = "JHEP",
      volume         = "06",
      pages          = "043",
      doi            = "10.1007/JHEP06(2010)043",
      year           = "2010",
      eprint         = "1002.2581",
      archivePrefix  = "arXiv",
      primaryClass   = "hep-ph",
}

@article{deFlorian:2016spz,
      author         = "de Florian, D. and others",
      title          = "{Handbook of LHC Higgs Cross Sections: 4. Deciphering the
                        Nature of the Higgs Sector}",
      collaboration  = "LHC Higgs Cross Section Working Group",
      doi            = "10.23731/CYRM-2017-002",
      year           = "2017",
      eprint         = "1610.07922",
      archivePrefix  = "arXiv",
      primaryClass   = "hep-ph",
      reportNumber   = "FERMILAB-FN-1025-T, CERN-2017-002-M",
      SLACcitation   = "%%CITATION = ARXIV:1610.07922;%%"
}

@Article{Gleisberg:2008fv,
      author         = "Gleisberg, Tanju and H{\"o}che, Stefan",
      title          = "{Comix, a new matrix element generator}",
      journal        = "JHEP",
      volume         = "12",
      pages          = "039",
      doi            = "10.1088/1126-6708/2008/12/039",
      year           = "2008",
      eprint         = "0808.3674",
      archivePrefix  = "arXiv",
      primaryClass   = "hep-ph",
}

@Article{Schumann:2007mg,
      author         = "Schumann, Steffen and Krauss, Frank",
      title          = "{A parton shower algorithm based on Catani--Seymour dipole factorisation}",
      journal        = "JHEP",
      volume         = "03",
      pages          = "038",
      doi            = "10.1088/1126-6708/2008/03/038",
      year           = "2008",
      eprint         = "0709.1027",
      archivePrefix  = "arXiv",
      primaryClass   = "hep-ph",
}

@article{Hoeche:2012yf,
      author         = "H{\"o}che, Stefan and Krauss, Frank and Sch{\"o}nherr, Marek and
                        Siegert, Frank",
      title          = "{QCD matrix elements + parton showers. The NLO case}",
      journal        = "JHEP",
      volume         = "04",
      year           = "2013",
      pages          = "027",
      doi            = "10.1007/JHEP04(2013)027",
      eprint         = "1207.5030",
      archivePrefix  = "arXiv",
      primaryClass   = "hep-ph",
      reportNumber   = "SLAC-PUB-15191, IPPP-12-52, DCPT-12-104, LPN12-081,
                        FR-PHENO-2012-017, MCNET-12-09, --FR-PHENO-2012-017",
      SLACcitation   = "%%CITATION = ARXIV:1207.5030;%%"
}

@article{Bothmann:2019yzt,
      author = "Bothmann, Enrico and others",
      title = "{Event generation with Sherpa 2.2}",
      journal = "SciPost Phys.",
      volume = "7",
      year = "2019",
      number = "3",
      pages = "034",
      doi = "10.21468/SciPostPhys.7.3.034",
      reportNumber = "FERMILAB-PUB-19-218-T, SLAC-PUB-17433, IPPP/19/42, MCNET-19-11",
      eprint = "1905.09127",
      archivePrefix = "arXiv",
      primaryClass = "hep-ph",
}

@article{Cascioli:2011va,
      author         = "Cascioli, Fabio and Maierh{\"o}fer, Philipp and Pozzorini, Stefano",
      title          = "{Scattering Amplitudes with Open Loops}",
      journal        = "Phys. Rev. Lett.",
      volume         = "108",
      year           = "2012",
      pages          = "111601",
      doi            = "10.1103/PhysRevLett.108.111601",
      eprint         = "1111.5206",
      archivePrefix  = "arXiv",
      primaryClass   = "hep-ph",
      reportNumber   = "ZU-TH-23-11, LPN11-66",
      SLACcitation   = "%%CITATION = ARXIV:1111.5206;%%"
}

@article{Hamilton:2013fea,
      author         = "Hamilton, Keith and Nason, Paolo and Re, Emanuele and
                        Zanderighi, Giulia",
      title          = "{NNLOPS simulation of Higgs boson production}",
      journal        = "JHEP",
      volume         = "10",
      year           = "2013",
      pages          = "222",
      doi            = "10.1007/JHEP10(2013)222",
      eprint         = "1309.0017",
      archivePrefix  = "arXiv",
      primaryClass   = "hep-ph",
      reportNumber   = "MCNET-13-11, CERN-PH-TH-2013-205, OUTP-13-18P",
      SLACcitation   = "%%CITATION = ARXIV:1309.0017;%%"
}

@article{Hamilton:2015nsa,
      author         = "Hamilton, Keith and Nason, Paolo and Zanderighi, Giulia",
      title          = "{Finite quark-mass effects in the NNLOPS POWHEG+MiNLO
                        Higgs generator}",
      journal        = "JHEP",
      volume         = "05",
      year           = "2015",
      pages          = "140",
      doi            = "10.1007/JHEP05(2015)140",
      eprint         = "1501.04637",
      archivePrefix  = "arXiv",
      primaryClass   = "hep-ph",
      reportNumber   = "CERN-PH-TH-2015-006",
      SLACcitation   = "%%CITATION = ARXIV:1501.04637;%%"
}

@article{Agostinelli:2002hh,
      author         = "Agostinelli, S. and others",
      title          = "{\textsc{Geant4} -- a simulation toolkit}",
      journal        = "Nucl. Instrum. Meth. A",
      volume         = "506",
      year           = "2003",
      pages          = "250",
      doi            = "10.1016/S0168-9002(03)01368-8",
      reportNumber   = "SLAC-PUB-9350, FERMILAB-PUB-03-339",
      SLACcitation   = "%%CITATION = NUIMA,A506,250;%%"
}

@Article{Read:2002hq,
    Author = {Read, Alexander L.},
    Journal = {J. Phys. G},
    Pages = {2693},
    doi = "10.1088/0954-3899/28/10/313",
    Title = "{Presentation of search results: the \(CL_S\) technique}",
    Volume = {28},
    Year = {2002}
}

@Article{PERF-2007-01,
    author         = "{ATLAS Collaboration}",
    title          = "{The ATLAS Experiment at the CERN Large Hadron Collider}",
    journal        = "JINST",
    volume         = "3",
    year           = "2008",
    pages          = "S08003",
    doi            = "10.1088/1748-0221/3/08/S08003",
    primaryClass   = "hep-ex",
}

@Article{SOFT-2010-01,
    author         = "{ATLAS Collaboration}",
    title          = "{The ATLAS Simulation Infrastructure}",
    journal        = "Eur. Phys. J. C",
    volume         = "70",
    year           = "2010",
    pages          = "823",
    doi            = "10.1140/epjc/s10052-010-1429-9",
    eprint         = "1005.4568",
    archivePrefix  = "arXiv",
    primaryClass   = "physics.ins-det",
}

@Article{EXOT-2013-15,
    author         = "{ATLAS Collaboration}",
    title          = "{Search for new light gauge bosons in Higgs boson decays to four-lepton final states in \(pp\) collisions at \(\sqrt{s} = 8\,\text{TeV}\) with the ATLAS detector at the LHC}",
    journal        = "Phys. Rev. D",
    volume         = "92",
    year           = "2015",
    pages          = "092001",
    doi            = "10.1103/PhysRevD.92.092001",
    reportNumber   = "CERN-PH-EP-2015-111",
    eprint         = "1505.07645",
    archivePrefix  = "arXiv",
    primaryClass   = "hep-ex",
}

@Article{HIGG-2014-02,
    author         = "{ATLAS Collaboration}",
    title          = "{Search for Higgs bosons decaying to \(aa\) in the \(\mu\mu\tau\tau\) final state in \(pp\) collisions at \(\sqrt{s} = 8\,\text{TeV}\) with the ATLAS experiment}",
    journal        = "Phys. Rev. D",
    volume         = "92",
    year           = "2015",
    pages          = "052002",
    doi            = "10.1103/PhysRevD.92.052002",
    reportNumber   = "CERN-PH-EP-2015-057",
    eprint         = "1505.01609",
    archivePrefix  = "arXiv",
    primaryClass   = "hep-ex",
}

@Article{PERF-2014-07,
    author         = "{ATLAS Collaboration}",
    title          = "{Topological cell clustering in the ATLAS calorimeters and its performance in LHC Run~1}",
    journal        = "Eur. Phys. J. C",
    volume         = "77",
    year           = "2017",
    pages          = "490",
    doi            = "10.1140/epjc/s10052-017-5004-5",
    reportNumber   = "CERN-PH-EP-2015-304",
    eprint         = "1603.02934",
    archivePrefix  = "arXiv",
    primaryClass   = "hep-ex",
}

@Article{PERF-2015-01,
    author         = "{ATLAS Collaboration}",
    title          = "{Reconstruction of primary vertices at the ATLAS experiment in Run~1 proton--proton collisions at the LHC}",
    journal        = "Eur. Phys. J. C",
    volume         = "77",
    year           = "2017",
    pages          = "332",
    doi            = "10.1140/epjc/s10052-017-4887-5",
    reportNumber   = "CERN-EP-2016-150",
    eprint         = "1611.10235",
    archivePrefix  = "arXiv",
    primaryClass   = "physics.ins-det",
}

@Article{EXOT-2016-22,
    author         = "{ATLAS Collaboration}",
    title          = "{Search for Higgs boson decays to beyond-the-Standard-Model light bosons in four-lepton events with the ATLAS detector at \(\sqrt{s} = 13\,\text{TeV}\)}",
    journal        = "JHEP",
    volume         = "06",
    year           = "2018",
    pages          = "166",
    doi            = "10.1007/JHEP06(2018)166",
    reportNumber   = "CERN-EP-2017-293",
    eprint         = "1802.03388",
    archivePrefix  = "arXiv",
    primaryClass   = "hep-ex",
}

@Article{HIGG-2016-03,
    author         = "{ATLAS Collaboration}",
    title          = "{Search for Higgs boson decays into a pair of light bosons in the \(bb\mu\mu\) final state in \(pp\) collision at \(\sqrt{s} = 13\,\text{TeV}\) with the ATLAS detector}",
    journal        = "Phys. Lett. B",
    volume         = "790",
    year           = "2019",
    pages          = "1",
    doi            = "10.1016/j.physletb.2018.10.073",
    reportNumber   = "CERN-EP-2018-153",
    eprint         = "1807.00539",
    archivePrefix  = "arXiv",
    primaryClass   = "hep-ex",
}

@Article{TRIG-2016-01,
    author         = "{ATLAS Collaboration}",
    title          = "{Performance of the ATLAS trigger system in 2015}",
    journal        = "Eur. Phys. J. C",
    volume         = "77",
    year           = "2017",
    pages          = "317",
    doi            = "10.1140/epjc/s10052-017-4852-3",
    reportNumber   = "CERN-EP-2016-241",
    eprint         = "1611.09661",
    archivePrefix  = "arXiv",
    primaryClass   = "hep-ex",
}

@Article{HIGG-2017-05,
    author         = "{ATLAS Collaboration}",
    title          = "{Search for the Higgs boson produced in association with a vector boson and decaying into two spin-zero particles in the \(H \rightarrow aa \rightarrow 4b\) channel in \(pp\) collisions at \(\sqrt{s} = 13\,\text{TeV}\) with the ATLAS detector}",
    journal        = "JHEP",
    volume         = "10",
    year           = "2018",
    pages          = "031",
    doi            = "10.1007/JHEP10(2018)031",
    reportNumber   = "CERN-EP-2018-128",
    eprint         = "1806.07355",
    archivePrefix  = "arXiv",
    primaryClass   = "hep-ex",
}

@Article{HIGG-2017-09,
    author         = "{ATLAS Collaboration}",
    title          = "{Search for Higgs boson decays into pairs of light (pseudo)scalar particles in the \(\gamma\gamma jj\) final state in \(pp\) collisions at \(\sqrt{s} = 13\,\text{TeV}\) with the ATLAS detector}",
    journal        = "Phys. Lett. B",
    volume         = "782",
    year           = "2018",
    pages          = "750",
    doi            = "10.1016/j.physletb.2018.06.011",
    reportNumber   = "CERN-EP-2017-295",
    eprint         = "1803.11145",
    archivePrefix  = "arXiv",
    primaryClass   = "hep-ex",
}

@Article{DAPR-2018-01,
    author         = "{ATLAS Collaboration}",
    title          = "{ATLAS data quality operations and performance for 2015--2018 data-taking}",
    journal        = "JINST",
    volume         = "15",
    year           = "2020",
    pages          = "P04003",
    doi            = "10.1088/1748-0221/15/04/P04003",
    reportNumber   = "CERN-EP-2019-207",
    eprint         = "1911.04632",
    archivePrefix  = "arXiv",
    primaryClass   = "physics.ins-det",
}

@Article{EGAM-2018-01,
    author         = "{ATLAS Collaboration}",
    title          = "{Electron and photon performance measurements with the ATLAS detector using the 2015--2017 LHC proton--proton collision data}",
    journal        = "JINST",
    volume         = "14",
    year           = "2019",
    pages          = "P12006",
    doi            = "10.1088/1748-0221/14/12/P12006",
    reportNumber   = "CERN-EP-2019-145",
    eprint         = "1908.00005",
    archivePrefix  = "arXiv",
    primaryClass   = "hep-ex",
}

@Article{HDBS-2018-47,
    author         = "{ATLAS Collaboration}",
    title          = "{Search for Higgs boson decays into two new low-mass spin-0 particles in the \(4b\) channel with the ATLAS detector using \(pp\) collisions at \(\sqrt{s} = 13\,\text{TeV}\)}",
    journal        = "Phys. Rev. D",
    volume         = "102",
    year           = "2020",
    pages          = "112006",
    doi            = "10.1103/PhysRevD.102.112006",
    reportNumber   = "CERN-EP-2020-067",
    eprint         = "2005.12236",
    archivePrefix  = "arXiv",
    primaryClass   = "hep-ex",
}

@Article{HDBS-2018-55,
    author         = "{ATLAS Collaboration}",
    title          = "{Search for Higgs bosons decaying into new spin-0 or spin-1 particles in four-lepton final states with the ATLAS detector with \(139\,\text{fb}^{-1}\) of \(pp\) collision data at \(\sqrt{s} = 13\,\text{TeV}\)}",
    journal        = "JHEP",
    volume         = "03",
    year           = "2022",
    pages          = "041",
    doi            = "10.1007/JHEP03(2022)041",
    reportNumber   = "CERN-EP-2021-193",
    eprint         = "2110.13673",
    archivePrefix  = "arXiv",
    primaryClass   = "hep-ex",
}

@Article{MUON-2018-03,
    author         = "{ATLAS Collaboration}",
    title          = "{Muon reconstruction and identification efficiency in ATLAS using the full Run~2 \(pp\) collision data set at \(\sqrt{s} = 13\,\text{TeV}\)}",
    journal        = "Eur. Phys. J. C",
    volume         = "81",
    year           = "2021",
    pages          = "578",
    doi            = "10.1140/epjc/s10052-021-09233-2",
    reportNumber   = "CERN-EP-2020-199",
    eprint         = "2012.00578",
    archivePrefix  = "arXiv",
    primaryClass   = "hep-ex",
}

@Article{TRIG-2018-01,
    author         = "{ATLAS Collaboration}",
    title          = "{Performance of the ATLAS muon triggers in Run~2}",
    journal        = "JINST",
    volume         = "15",
    year           = "2020",
    pages          = "P09015",
    doi            = "10.1088/1748-0221/15/09/p09015",
    reportNumber   = "CERN-EP-2020-031",
    eprint         = "2004.13447",
    archivePrefix  = "arXiv",
    primaryClass   = "physics.ins-det",
}

@Article{TRIG-2018-05,
    author         = "{ATLAS Collaboration}",
    title          = "{Performance of electron and photon triggers in ATLAS during LHC Run~2}",
    journal        = "Eur. Phys. J. C",
    volume         = "80",
    year           = "2020",
    pages          = "47",
    doi            = "10.1140/epjc/s10052-019-7500-2",
    reportNumber   = "CERN-EP-2019-169",
    eprint         = "1909.00761",
    archivePrefix  = "arXiv",
    primaryClass   = "hep-ex",
}

@Article{EGAM-2019-01,
    author         = "{ATLAS Collaboration}",
    title          = "{Tools for estimating fake/non-prompt lepton backgrounds with the ATLAS detector at the LHC}",
    journal        = "JINST",
    volume         = "18",
    year           = "2023",
    pages          = "T11004",
    doi            = "10.1088/1748-0221/18/11/T11004",
    reportNumber   = "CERN-EP-2022-214",
    eprint         = "2211.16178",
    archivePrefix  = "arXiv",
    primaryClass   = "hep-ex",
}

@Article{TRIG-2019-03,
    author         = "{ATLAS Collaboration}",
    title          = "{The ATLAS inner detector trigger performance in \(pp\) collisions at \(13\,\text{TeV}\) during LHC Run~2}",
    journal        = "Eur. Phys. J. C",
    volume         = "82",
    year           = "2022",
    pages          = "206",
    doi            = "10.1140/epjc/s10052-021-09920-0",
    reportNumber   = "CERN-EP-2021-076",
    eprint         = "2107.02485",
    archivePrefix  = "arXiv",
    primaryClass   = "hep-ex",
}

@Article{HDBS-2021-13,
    author         = "{ATLAS Collaboration}",
    title          = "{Search for a new scalar decaying into new spin-1 bosons in four-lepton final states with the ATLAS detector}",
    year           = "2024",
    reportNumber   = "CERN-EP-2024-248",
    eprint         = "2410.16781",
    archivePrefix  = "arXiv",
    primaryClass   = "hep-ex",
}

@Article{HIGG-2021-05,
    author         = "{ATLAS Collaboration}",
    title          = "{Combination of searches for invisible decays of the Higgs boson using \(139\,\text{fb}^{-1}\) of proton--proton collision data at \(\sqrt{s} = 13\,\text{TeV}\) collected with the ATLAS experiment}",
    journal        = "Phys. Lett. B",
    volume         = "842",
    year           = "2023",
    pages          = "137963",
    doi            = "10.1016/j.physletb.2023.137963",
    reportNumber   = "CERN-EP-2022-289",
    eprint         = "2301.10731",
    archivePrefix  = "arXiv",
    primaryClass   = "hep-ex",
}

@Article{SOFT-2022-02,
    author         = "{ATLAS Collaboration}",
    title          = "{Software and computing for Run~3 of the ATLAS experiment at the LHC}",
    year           = "2024",
    reportNumber   = "CERN-EP-2024-100",
    eprint         = "2404.06335",
    archivePrefix  = "arXiv",
    primaryClass   = "hep-ex",
}

@Booklet{ATL-SOFT-PUB-2025-001,
    author         = "{ATLAS Collaboration}",
    title          = "{ATLAS Computing Acknowledgements}",
    howpublished   = "{ATL-SOFT-PUB-2025-001}",
    url            = "https://cds.cern.ch/record/2922210",
    year           = 2025
}

@Article{LUCID2,
    author         = "Avoni, G. and others",
    title          = "{The new LUCID-2 detector for luminosity measurement and monitoring in ATLAS}",
    journal        = "JINST",
    volume         = "13",
    year           = "2018",
    pages          = "P07017",
    doi            = "10.1088/1748-0221/13/07/P07017",
}

@Article{CMS-HIG-13-010,
    author         = "{CMS Collaboration}",
    title          = "{A search for pair production of new light bosons decaying into muons}",
    journal        = "Phys. Lett. B",
    volume         = "752",
    year           = "2016",
    pages          = "146",
    doi            = "10.1016/j.physletb.2015.10.067",
    reportNumber   = "CERN-PH-EP-2015-116",
    eprint         = "1506.00424",
    archivePrefix  = "arXiv",
    primaryClass   = "hep-ex",
}

@Article{CMS-HIG-16-015,
    author         = "{CMS Collaboration}",
    title          = "{Search for light bosons in decays of the \(125\,\text{GeV}\) Higgs boson in proton--proton collisions at \(\sqrt{s} = 8\,\text{TeV}\)}",
    journal        = "JHEP",
    volume         = "10",
    year           = "2017",
    pages          = "076",
    doi            = "10.1007/JHEP10(2017)076",
    reportNumber   = "CERN-EP-2016-292",
    eprint         = "1701.02032",
    archivePrefix  = "arXiv",
    primaryClass   = "hep-ex",
}

@Article{CMS-HIG-18-024,
    author         = "{CMS Collaboration}",
    title          = "{Search for a light pseudoscalar Higgs boson in the boosted \(\mu\mu\tau\tau\) final state in proton--proton collisions at \(\sqrt{s} = 13\,\text{TeV}\)}",
    journal        = "JHEP",
    volume         = "08",
    year           = "2020",
    pages          = "139",
    doi            = "10.1007/JHEP08(2020)139",
    reportNumber   = "CERN-EP-2020-061",
    eprint         = "2005.08694",
    archivePrefix  = "arXiv",
    primaryClass   = "hep-ex",
}

@Article{CMS-EXO-19-018,
    author         = "{CMS Collaboration}",
    title          = "{Search for a Narrow Resonance Lighter than \(200\,\text{GeV}\) Decaying to a Pair of Muons in Proton--Proton Collisions at \(\sqrt{s} = 13\,\text{TeV}\)}",
    journal        = "Phys. Rev. Lett.",
    volume         = "124",
    year           = "2020",
    pages          = "131802",
    doi            = "10.1103/PhysRevLett.124.131802",
    reportNumber   = "CERN-EP-2019-265",
    eprint         = "1912.04776",
    archivePrefix  = "arXiv",
    primaryClass   = "hep-ex",
}

@Article{CMS-HIG-19-007,
    author         = "{CMS Collaboration}",
    title          = "{Search for low-mass dilepton resonances in Higgs boson decays to four-lepton final states in proton--proton collisions at \(\sqrt{s} = 13\,\text{TeV}\)}",
    journal        = "Eur. Phys. J. C",
    volume         = "82",
    year           = "2022",
    pages          = "290",
    doi            = "10.1140/epjc/s10052-022-10127-0",
    reportNumber   = "CERN-EP-2021-093",
    eprint         = "2111.01299",
    archivePrefix  = "arXiv",
    primaryClass   = "hep-ex",
}

@Article{CMS-EXO-20-014,
    author         = "{CMS Collaboration}",
    title          = "{Search for long-lived particles decaying into muon pairs in proton--proton collisions at \(\sqrt{s}\ = 13\,\text{TeV}\) collected with a dedicated high-rate data stream}",
    journal        = "JHEP",
    volume         = "04",
    year           = "2022",
    pages          = "062",
    doi            = "10.1007/JHEP04(2022)062",
    reportNumber   = "CERN-EP-2021-266",
    eprint         = "2112.13769",
    archivePrefix  = "arXiv",
    primaryClass   = "hep-ex",
}

@Article{CMS-HIG-20-003,
    author         = "{CMS Collaboration}",
    title          = "{Search for invisible decays of the Higgs boson produced via vector boson fusion in proton--proton collisions at \(\sqrt{s} = 13\,\text{TeV}\)}",
    journal        = "Phys. Rev. D",
    volume         = "105",
    year           = "2022",
    pages          = "092007",
    doi            = "10.1103/PhysRevD.105.092007",
    reportNumber   = "CERN-EP-2021-273",
    eprint         = "2201.11585",
    archivePrefix  = "arXiv",
    primaryClass   = "hep-ex",
}

@Article{CMS-EXO-23-005,
    author         = "{CMS Collaboration}",
    title          = "{Dark sector searches with the CMS experiment}",
    year           = "2024",
    reportNumber   = "CERN-EP-2024-106",
    eprint         = "2405.13778",
    archivePrefix  = "arXiv",
    primaryClass   = "hep-ex",
}

@Booklet{ATLAS-CONF-2015-029,
    author         = "{ATLAS Collaboration}",
    title          = "{Selection of jets produced in \(13~\text{TeV}\) proton--proton collisions with the ATLAS detector}",
    howpublished   = "{ATLAS-CONF-2015-029}",
    url            = "https://cds.cern.ch/record/2037702",
    year           = "2015",
}

@Booklet{ATLAS-CONF-2019-021,
    author         = "{ATLAS Collaboration}",
    title          = "{Luminosity determination in \(pp\) collisions at \(\sqrt{s} = 13\,\text{TeV}\) using the ATLAS detector at the LHC}",
    howpublished   = "{ATLAS-CONF-2019-021}",
    url            = "https://cds.cern.ch/record/2677054",
    year           = "2019",
}

@Booklet{ATL-PHYS-PUB-2010-013,
    author         = "{ATLAS Collaboration}",
    title          = "{The simulation principle and performance of the ATLAS fast calorimeter simulation FastCaloSim}",
    howpublished   = "{ATL-PHYS-PUB-2010-013}",
    url            = "https://cds.cern.ch/record/1300517",
    year           = "2010",
}

@Booklet{ATL-PHYS-PUB-2014-021,
    author         = "{ATLAS Collaboration}",
    title          = "{ATLAS Pythia~8 tunes to \(7~\text{TeV}\) data}",
    howpublished   = "{ATL-PHYS-PUB-2014-021}",
    url            = "https://cds.cern.ch/record/1966419",
    year           = "2014",
}

@Booklet{ATL-PHYS-PUB-2017-005,
    author         = "{ATLAS Collaboration}",
    title          = "{Multi-Boson Simulation for \(13~\text{TeV}\) ATLAS Analyses}",
    howpublished   = "{ATL-PHYS-PUB-2017-005}",
    url            = "https://cds.cern.ch/record/2261933",
    year           = "2017",
}

@Article{wells2008find,
  author        = {Wells, James D.},
  title         = {{How to Find a Hidden World at the Large Hadron Collider}},
  year          = {2008},
  archiveprefix = {arXiv},
  eprint        = {0803.1243},
  primaryclass  = {hep-ph},
  reportnumber  = {MCTP-07-51, CERN-PH-TH-2008-47},
  slaccitation  = {%%CITATION = ARXIV:0803.1243;%%},
}

@Article{Caola:2015psa,
  author        = {Caola, Fabrizio and Melnikov, Kirill and Röntsch, Raoul and Tancredi, Lorenzo},
  title         = {{QCD corrections to ZZ production in gluon fusion at the LHC}},
  journal       = {Phys. Rev. D},
  year          = {2015},
  volume        = {92},
  number        = {9},
  pages         = {094028},
  archiveprefix = {arXiv},
  doi           = {10.1103/PhysRevD.92.094028},
  eprint        = {1509.06734},
  primaryclass  = {hep-ph},
  reportnumber  = {TTP15-035, CERN-PH-TH-2015-226, FERMILAB-PUB-15-398-T},
  slaccitation  = {%%CITATION = ARXIV:1509.06734;%%},
}

@TechReport{cranmer2012histfactory,
  author        = {Cranmer, Kyle and Shibata, Akira and Verkerke, Wouter and Moneta, Lorenzo and Lewis, George},
  title         = {HistFactory: A tool for creating statistical models for use with RooFit and RooStats},
  year          = {2012},
  reportnumber  = {CERN-OPEN-2012-016},
  url           = {https://cds.cern.ch/record/1456844},
  __markedentry = {[Chris:]},
}

@Article{Moneta:2010pm,
  author        = {Moneta, Lorenzo and Belasco, Kevin and Cranmer, Kyle S. and Kreiss, S. and Lazzaro, Alfio and Piparo, Danilo and Schott, Gregory and Verkerke, Wouter and Wolf, Matthias},
  title         = {{The RooStats Project}},
  journal       = {PoS},
  year          = {2010},
  volume        = {ACAT2010},
  pages         = {057},
  __markedentry = {[Chris:]},
  archiveprefix = {arXiv},
  booktitle     = {{Proceedings, 13th International Workshop on Advanced computing and analysis techniques in physics research (ACAT2010): Jaipur, India, February 22-27, 2010}},
  doi           = {10.22323/1.093.0057},
  eprint        = {1009.1003},
  primaryclass  = {physics.data-an},
  slaccitation  = {%%CITATION = ARXIV:1009.1003;%%},
}

@article{Contino:2014aaa,
    author = "Contino, Roberto and Ghezzi, Margherita and Grojean, Christophe and Mühlleitner, Margarete and Spira, Michael",
    title = "{eHDECAY: an implementation of the Higgs effective Lagrangian into HDECAY}",
    eprint = "1403.3381",
    archivePrefix = "arXiv",
    primaryClass = "hep-ph",
    reportNumber = "CERN-PH-TH-2013-316, KA-TP-03-2014, PSI-PR-13-20",
    doi = "10.1016/j.cpc.2014.06.028",
    journal = "Comput. Phys. Commun.",
    volume = "185",
    pages = "3412--3423",
    year = "2014"
}

@Article{Curtin:2013fra,
  author        = {Curtin, David and Essig, Rouven and Gori, Stefania and Jaiswal, Prerit and Katz, Andrey and others},
  title         = {{Exotic decays of the 125 GeV Higgs boson}},
  journal       = {\PRD{}},
  year          = {2014},
  volume        = {90},
  number        = {7},
  pages         = {075004},
  archiveprefix = {arXiv},
  doi           = {10.1103/PhysRevD.90.075004},
  eprint        = {1312.4992},
  primaryclass  = {hep-ph},
  reportnumber  = {YITP-13-47, PITT-PACC-1314},
  slaccitation  = {%%CITATION = ARXIV:1312.4992;%%},
}

@Article{Davoudiasl:2013aya,
  author        = {Davoudiasl, Hooman and Lee, Hye-Sung and Lewis, Ian and Marciano, William J.},
  title         = {{Higgs decays as a window into the dark sector}},
  journal       = {\PRD{}},
  year          = {2013},
  volume        = {88},
  number        = {1},
  pages         = {015022},
  archiveprefix = {arXiv},
  doi           = {10.1103/PhysRevD.88.015022},
  eprint        = {1304.4935},
  primaryclass  = {hep-ph},
  publisher     = {APS},
  reportnumber  = {JLAB-THY-14-1844},
  slaccitation  = {%%CITATION = ARXIV:1304.4935;%%},
}

@Article{gopalakrishna2008higgs,
  author        = {Gopalakrishna, Shrihari and Jung, Sunghoon and Wells, James D.},
  title         = {{Higgs boson decays to four fermions through an Abelian hidden sector}},
  journal       = {Phys. Rev. D},
  year          = {2008},
  volume        = {78},
  pages         = {055002},
  archiveprefix = {arXiv},
  doi           = {10.1103/PhysRevD.78.055002},
  eprint        = {0801.3456},
  primaryclass  = {hep-ph},
  reportnumber  = {CERN-PH-TH-2008-11, MCTP-07-47, BNL-HET-08-2},
  slaccitation  = {%%CITATION = ARXIV:0801.3456;%%},
}

@Article{LHCHiggsCrossSectionWorkingGroup:2011ti,
  author        = {Dittmaier, S. and others},
  title         = {{Handbook of LHC Higgs cross sections: 1. Inclusive observables}},
  journal       = {CERN-2011-002},
  year          = {2011},
  archiveprefix = {arXiv},
  collaboration = {LHC Higgs Cross Section Working Group},
  doi           = {10.5170/CERN-2011-002},
  eprint        = {1101.0593},
  primaryclass  = {hep-ph},
  reportnumber  = {CERN-2011-002},
  slaccitation  = {%%CITATION = ARXIV:1101.0593;%%},
}

@Article{Cowan:2010js-witherratum,
   Author = {Cowan, Glen and Cranmer, Kyle and Gross, Eilam and Vitells, Ofer},
   Title = {{Asymptotic formulae for likelihood-based tests of new physics}},
   Journal = {Eur. Phys. J. C},
   Volume = {71},
   Year = {2011},
   Pages = {1554},
   doi = "10.1140/epjc/s10052-011-1554-0",
   Eprint = "1007.1727",
   Archiveprefix = {arXiv},
   Primaryclass = {physics.data-an},
   addendum = "Erratum: \href{http://dx.doi.org/10.1140/epjc/s10052-013-2501-z}{Eur. Phys. J. C \textbf{73} (2013) 2501}",
}
\clearpage
 
\begin{flushleft}
\hypersetup{urlcolor=black}
{\Large The ATLAS Collaboration}

\bigskip

\AtlasOrcid[0000-0002-6665-4934]{G.~Aad}$^\textrm{\scriptsize 102}$,
\AtlasOrcid[0000-0002-5888-2734]{B.~Abbott}$^\textrm{\scriptsize 120}$,
\AtlasOrcid[0000-0002-1002-1652]{K.~Abeling}$^\textrm{\scriptsize 55}$,
\AtlasOrcid[0000-0001-5763-2760]{N.J.~Abicht}$^\textrm{\scriptsize 49}$,
\AtlasOrcid[0000-0002-8496-9294]{S.H.~Abidi}$^\textrm{\scriptsize 29}$,
\AtlasOrcid[0000-0002-9987-2292]{A.~Aboulhorma}$^\textrm{\scriptsize 35e}$,
\AtlasOrcid[0000-0001-5329-6640]{H.~Abramowicz}$^\textrm{\scriptsize 152}$,
\AtlasOrcid[0000-0002-1599-2896]{H.~Abreu}$^\textrm{\scriptsize 151}$,
\AtlasOrcid[0000-0003-0403-3697]{Y.~Abulaiti}$^\textrm{\scriptsize 117}$,
\AtlasOrcid[0000-0003-0762-7204]{A.C.~Abusleme~Hoffman}$^\textrm{\scriptsize 137a}$
% ...
\bigskip
\\

% ...
$^{29}$Physics Department, Brookhaven National Laboratory, Upton NY; United States of America.\\
% ...
$^{35}$$^{(a)}$Facult\'e des Sciences Ain Chock, Universit\'e Hassan II de Casablanca;$^{(b)}$Facult\'{e} des Sciences, Universit\'{e} Ibn-Tofail, K\'{e}nitra;$^{(c)}$Facult\'e des Sciences Semlalia, Universit\'e Cadi Ayyad, LPHEA-Marrakech;$^{(d)}$LPMR, Facult\'e des Sciences, Universit\'e Mohamed Premier, Oujda;$^{(e)}$Facult\'e des sciences, Universit\'e Mohammed V, Rabat;$^{(f)}$Institute of Applied Physics, Mohammed VI Polytechnic University, Ben Guerir; Morocco.\\
% ...
$^{49}$Fakult\"{a}t Physik , Technische Universit{\"a}t Dortmund, Dortmund; Germany.\\
% ...
$^{55}$II. Physikalisches Institut, Georg-August-Universit\"{a}t G\"ottingen, G\"ottingen; Germany.\\
% ...
$^{102}$CPPM, Aix-Marseille Universit\'e, CNRS/IN2P3, Marseille; France.\\
% ...
$^{117}$Department of Physics, New York University, New York NY; United States of America.\\
% ...
$^{120}$Homer L. Dodge Department of Physics and Astronomy, University of Oklahoma, Norman OK; United States of America.\\
% ...
$^{137}$$^{(a)}$Departamento de F\'isica, Pontificia Universidad Cat\'olica de Chile, Santiago;$^{(b)}$Millennium Institute for Subatomic physics at high energy frontier (SAPHIR), Santiago;$^{(c)}$Instituto de Investigaci\'on Multidisciplinario en Ciencia y Tecnolog\'ia, y Departamento de F\'isica, Universidad de La Serena;$^{(d)}$Universidad Andres Bello, Department of Physics, Santiago;$^{(e)}$Instituto de Alta Investigaci\'on, Universidad de Tarapac\'a, Arica;$^{(f)}$Departamento de F\'isica, Universidad T\'ecnica Federico Santa Mar\'ia, Valpara\'iso; Chile.\\
% ...
$^{151}$Department of Physics, Technion, Israel Institute of Technology, Haifa; Israel.\\
$^{152}$Raymond and Beverly Sackler School of Physics and Astronomy, Tel Aviv University, Tel Aviv; Israel.\\
% ...
$^{a}$ Also Affiliated with an institute covered by a cooperation agreement with CERN.\\
$^{b}$ Also at An-Najah National University, Nablus; Palestine.\\
$^{c}$ Also at Borough of Manhattan Community College, City University of New York, New York NY; United States of America.\\
$^{d}$ Also at Center for High Energy Physics, Peking University; China.\\
$^{e}$ Also at Center for Interdisciplinary Research and Innovation (CIRI-AUTH), Thessaloniki; Greece.\\
$^{f}$ Also at Centro Studi e Ricerche Enrico Fermi; Italy.\\
% ...

\end{flushleft}


%

%
%

%

%
%
%
%

\end{document}